\documentclass[12pt,draftcls,onecolumn]{IEEEtran}

\ifCLASSINFOpdf
\else
\fi
\usepackage{graphics} 
\usepackage{epsfig} 
\usepackage{epstopdf}
\usepackage{amsmath} 
\usepackage{amssymb}  

\usepackage{cite}
\usepackage{graphicx}
\usepackage{pgf}
\usepackage{bbm}
\usepackage{etoolbox}
\usepackage{mathrsfs}
\usepackage{enumerate}
\usepackage{color}
\usepackage[textsize=footnotesize,textwidth=1.5cm]{todonotes}	

\newtheorem{thm}{Theorem}
\newtheorem{assumption}{Assumption}
\newtheorem{lem}[thm]{Lemma}
\newtheorem{definition}{Definition}
\newtheorem{rem}{Remark}

\newcommand{\jln}[1]{\textcolor{black}{#1}}

\newcommand{\rsa}[1]{\textcolor{black}{#1}}
\newtoggle{jou}
\togglefalse{jou}

\begin{document}
%
\title{A Dynamic Game Model of Collective Choice\\ in Multi-Agent Systems}
%
%
%

\author{Rabih~Salhab,
        Roland~P.~Malham\'e               
        and~Jerome~Le~Ny
\thanks{This work was supported by NSERC under Grants 6820-2011 and 435905-13.
The authors are with the department of Electrical Engineering, 
Polytechnique Montreal and with GERAD, Montreal, QC H3T-1J4, Canada {\tt\small \{rabih.salhab, roland.malhame, jerome.le-ny\}@polymtl.ca}.

\jln{Preliminary versions of this paper appeared at CDC 2014 and CDC 2015 \cite{Rab2014, Rab2015}.}
}}

%
%

\markboth{}%
{Rabih \MakeLowercase{\textit{et al.}}: Bare Demo of IEEEtran.cls for Journals}
%



\maketitle

\begin{abstract}
Inspired by successful biological collective decision mechanisms such as 
honey bees searching for a new colony or the collective navigation of 
fish schools, we consider a mean field games (MFG)-like scenario 
where a large number of agents have to make a choice among a set of
different potential target destinations. Each individual both influences
and is influenced by the group's decision, as represented by the mean 
trajectory of all agents.
The model can be interpreted as a stylized version of opinion crystallization 
in an election for example.
The agents' biases are dictated first by their initial spatial position and, in a subsequent 
generalization of the model, by a combination of initial position and 
a priori individual preference. The agents have linear dynamics and are 
coupled through a modified form of quadratic cost. Fixed point based finite 
population  equilibrium conditions are identified and associated existence 
conditions are established. In general multiple equilibria may exist and
the agents need to know all initial conditions to compute them precisely. 
However, as the number of agents increases sufficiently, we show that (i) the computed 
fixed point equilibria qualify as epsilon Nash equilibria, 
(ii) agents no longer require all initial conditions to compute the equilibria 
but rather can do so based on a representative  probability distribution of 
these conditions now viewed as random variables. Numerical results are reported. 
\end{abstract}


%
\IEEEpeerreviewmaketitle

\section{Introduction}


\rsa{Collective decision making is a common phenomenon in social structures
ranging from animal populations \cite{Leonard12_decision, Couzin05_leadership}
to human societies \cite{Merill99_vote}.
Examples include honey bees searching for a new colony \cite{Seeley91_honeyBees}, 
\cite{Camazin99_honeyBees}, the navigation of fish schools \cite{Tien04_fishSchools}, 
\cite{Aoki_fishSimulation}, or quorum sensing \cite{Pratt06_algoDecisionMaking}.} 
\rsa{
Collective decisions involve dynamic ``microscopic-macroscopic" or ``individual-social''
interactions.
On the one hand, individual choices are socially influenced, that is influenced by the
behavior of the group. On the other hand, the collective behavior itself results 
from aggregating individual choices. 
In elections for example, 
an interplay between individual interests and collective opinion swings leads 
to the crystallization of final decisions \cite{Daron13_op, Hegselmann02_op, Merill99_vote}.}

``Homing'' optimal control problems, first introduced by Whittle and Gait 
in \cite{Whittle70_homing} and studied later in \cite{Whittle82, Kuhn85_homing, 
Lef04, Lef_GLQG} for example, are concerned with a single agent 
trying to reach one of multiple predefined final states. 
Here we consider a similar fundamental issue but in a multi-agent setting.
A large number of agents initially spread out in $\mathbb R^n$ need to move within 
a finite time horizon to one of multiple possible home or target destinations. They 
must do so while trying to remain tightly grouped and expending as little 
control effort as possible.
Our goal is to model situations in which the choice made by each agent regarding 
which destination to reach both influences and depends on the behavior of the population.
For example, when honey bees determine their next site to establish a colony, 
they must make a choice between different alternatives based on the information provided 
by scouts, who are themselves part of the group.
Even though certain colonies can be easier to reach and are more attractive for some bees,
following the majority is still a priority to enhance the foraging ability. 
%
Similarly, in a  navigation situation for  a collection of micro robots exploring 
an unknown terrain, remaining grouped may be necessary for achieving coordinated 
collective tasks \cite{Nourian11_collective, Mesbah10_graphth, DistCtrlRobotNetw, jerome13_ada}. 
In animal collective navigation, discrete choices must be made regarding the route
to take, but at the same time staying with the group offers better protection 
against predators. Finally, our model may be an abstract representation of 
opinion crystallization in an election where (i) relative distances measure current 
differences of opinions, (ii) individuals are sensitive to collective opinion swings, 
and (iii) a choice must be made before a finite deadline\cite{Daron13_op, Hegselmann02_op, 
Merill99_vote}.

 A related topic in economics is discrete choice models where an agent makes a choice between 
 multiple alternatives such as mode of transportation \cite{Kappelman2005}, entry and withdrawal 
 from the labor market, residential location \cite{Bhat2004}, or a physician \cite{burke2003physician}. 
 In many circumstances, these individual choices are influenced by the so called ``Peer Effect", 
 ``Neighborhood Effect" or ``Social Effect". 
 In particular, Brock and Durlauf \cite{Brock2001} use an approach similar to \jln{Mean Fied Games (MFG)
 \cite{Huang07_Large, Lasry07_MFG}}
 and inspired by statistical mechanics to study a static binary discrete choice model with a 
 large number of agents which takes into account the effect of the agents' interdependence on the 
 individual choices.
In their model, the individual choices are influenced by the mean of the other agents' choices, 
while for an infinite size population, the impact of an isolated individual choice on this mean 
is negligible. The authors show that in an infinite size rational population, each agent can 
predict this mean as the result of a fixed point calculation, and makes a decentralized choice 
based upon its prediction. Moreover, multiple anticipated means may exist. Our analysis leads 
to similar insights for a dynamic non-cooperative multiple choice game including situations 
where the agents have limited information about the dynamics of other agents. 
\jln{\section{Problem Statement and Contributions}}

\jln{In this section, we formulate our problem, state our main contributions and provide
an outline for the rest of the paper.}



\subsection{Deterministic Case}
 We consider a dynamic non-cooperative game involving $N$ players with identical linear dynamics
 \begin{align}		\label{eq: agents dynamics}
\dot{x}_i =Ax_{i}+Bu_i
&& \forall i \in \{ 1 \ldots, N\},   
\end{align}
where $x_i \in \mathbb{R}^n$ is the state of agent $i$ and $u_i \in \mathbb R^m$ its control input.
Player $i$, $1 \leq i \leq N$, is associated with an individual cost functional 
\iftoggle{jou}{
\begin{multline}	\label{eq:sysinde}
J_{i}(u_{i},\bar{x},x_i^0) =  \int_0^T \; \left \{ \frac{q}{2} \| x_{i} - \bar{x} \|^{2} + \frac{r}{2} \| u_{i} \|^{2} \right \} \,\mathrm{d}t \\ 
+ \frac{M}{2} \min\limits_{j=1,\dots,l} \Big(  \| x_{i}(T)-p_{j} \|^{2}   \Big),
\end{multline}}{
\begin{equation}	\label{eq:sysinde}
J_{i}(u_{i},\bar{x},x_i^0) =  \int_0^T \; \left \{ \frac{q}{2} \| x_{i} - \bar{x} \|^{2} + \frac{r}{2} \| u_{i} \|^{2} \right \} \,\mathrm{d}t 
+ \frac{M}{2} \min\limits_{j=1,\dots,l} \Big(  \| x_{i}(T)-p_{j} \|^{2}   \Big),
\end{equation}
}  
where $\bar{x}(t)  \triangleq 1/N \sum_{i=1}^N x_i(t)$, $p_{j} \in \mathbb R^n$ (for $j=1,\dots,l$), 
$q$, $r$ are positive constants and $M$ is a large positive number. The running cost requires the agents 
to develop as little effort as possible while moving and to stay grouped around the mean of the population. 
Moreover, each agent should reach before the final time $T$ one of the destinations $p_{j},j=1,\dots,l$.  
Otherwise, it is strongly penalized by the terminal cost. 
\jln{Hence, the overall individual cost captures the problem faced by each agent of deciding between 
a finite set of alternatives, while trying to remain close to the mean population trajectory.}
It is sometimes convenient to write the costs in a game theoretic form, i.e. $J_i(u_i,u_{-i})$, where $u_{-i}=(u_1,\dots,u_{i-1},u_{i+1},\dots,u_N)$. We seek $\epsilon$-Nash strategies, i.e. such that an agent can benefit at most $\epsilon$ through unilateral deviant behavior, with $\epsilon$ going to zero as $N$ goes to infinity \cite{Huang03_thesis}. We assume that each agent can observe only its own state and the initial states of the other agents. 
\begin{definition}
Consider $N$ players, a set of strategy profiles $S=S_1 \times \dots \times S_N$ and for each player $k$, a payoff function $J_{k}(u_1,\dots,u_N)$, $\forall (u_1,\dots,u_N) \in S$. A strategy profile $(u_1^*,\dots,u_N^*)\in S$ is called an $\epsilon-$Nash equilibrium with respect to the costs $J_{k}$ if there exists an $\epsilon>0$ such that for any fixed $1\leq i \leq N$ and for all $u_{i}\in S_i$, we have
\begin{align*}
J_i(u_i,u^*_{-i})\geq J_i(u_i^*,u_{-i}^*)-\epsilon.
\end{align*}
\end{definition}

Inspired by the framework of MFG theory 
\cite{Nourian11_collective, Huang03_thesis, Huang03_wirelessPower, Huang06_particles, Lasry07_MFG, Huang07_Large}
discussed in Section \ref{section: contributions} below, we develop a class of decentralized strategies based on a fixed point requirement. Identification of the strategies requires only  that an  agent  knows its own state and the initial states of the other agents. As we later show in the paper, when the number of agents $N$ increases without bound, these fixed point based strategies achieve their meaning as $\epsilon-$Nash equilibria.


\subsection{Stochastic Case}
As $N$ goes to infinity, it is also convenient to think of the initial states  as realizations of random variables resulting from a common probability distribution function in a collection of independent experiments. 
Agent $i$, for $1 \leq i \leq N$, is then associated with the following adequately modified cost:
\iftoggle{jou}{
\begin{multline}	\label{eq:sysin}
J_{i}(u_{i},\bar{x},x_i^0) =  \mathbb{E} \bigg ( \int_0^T \; \left \{ \frac{q}{2} \| x_{i} - \bar{x} \|^{2} + \frac{r}{2} \| u_{i} \|^{2} \right \} \,\mathrm{d}t \\ 
+ \frac{M}{2} \min\limits_{j=1,\dots,l} \Big(  \| x_{i}(T)-p_{j} \|^{2}   \Big) \Big | x_i^0 \bigg ).               
\end{multline}}{
\begin{equation}	\label{eq:sysin}
J_{i}(u_{i},\bar{x},x_i^0) =  \mathbb{E} \bigg ( \int_0^T \; \left \{ \frac{q}{2} \| x_{i} - \bar{x} \|^{2} + \frac{r}{2} \| u_{i} \|^{2} \right \} \,\mathrm{d}t 
+ \frac{M}{2} \min\limits_{j=1,\dots,l} \Big(  \| x_{i}(T)-p_{j} \|^{2}   \Big) \Big | x_i^0 \bigg ).               
\end{equation}}
In this case, we establish that an agent only needs to know its own state and the common initial states probability distribution to construct  one of the decentralized fixed point based strategies alluded to earlier.

\subsection{The MFG Approach and our Contributions}	\label{section: contributions}

\jln{The MFG} \rsa{approach is concerned with a class of dynamic non-cooperative games involving 
a large number of players where the individual strategies are considerably affected by the 
mass behavior, while the influence of an isolated individual strategy on the group is negligible. 
Linear Quadratic Gaussian (LQG) MFG problems were developed in \cite{Huang03_thesis, Huang03_wirelessPower, Huang07_Large}, 
while the general nonlinear stochastic framework was considered 
in \cite{huang2006large, lasry2006jeux, lasry2006jeux2, Lasry07_MFG}.} 
The MFG approach posits at the outset an infinite population to which one can ascribe a deterministic although initially unknown macroscopic behavior. Hence, one starts by assuming that the mean field contributed term 
$\bar{x}$ in the cost is given and equal to  some $\hat{x}$. The cost functions being now decoupled, each agent optimally tracks $\hat{x}$. The resulting control laws are decentralized. This analysis of the tracking problem is presented in Section \ref{section: tracking}. 
By implementing the resulting decentralized strategies in the dynamics of the agents,  a new candidate tracking path is obtained by computing the corresponding  mean population trajectory. Indeed, and it is a fundamental argument in MFG analysis, asymptotically as the population grows, the posited  tracked path is an acceptable candidate only if it is reproduced as the mean of the agents when they optimally respond to it. Thus, we look for candidate trajectories which are fixed points of the tracking path to tracking path map  defined above.   
In Section \ref{section: fixed}, these 
fixed points are studied for the deterministic case with a finite population, 
and an explicit expression is obtained by assuming that each agent knows the exact initial states of all other agents. 
The alternative probabilistic description of the agents' initial states is explored in  Section \ref{section: fixedpointstoch}.  
In Section \ref{section: initialpreferences}, we further  generalize the problem formulation to include initial preferences towards the target destinations. Moreover, 
we consider that the agents have nonuniform dynamics and that each agent has limited 
information about the other agents dynamic parameters in the form of a statistical distribution over 
the matrices $A$ and $B$. 
Section \ref{section:Nash} shows that the decentralized strategies developed 
when tracking the fixed point trajectories constitute $\epsilon-$Nash equilibria
in all the cases considered above, with $\epsilon$ going to zero as $N$ goes to infinity. 
In Section \ref{section: simulation}, we provide some numerical simulation results, 
while Section \ref{section: conclusion} presents our conclusions.


\jln{Although we rely on the MFG methodology in order to analyze the behavior of many agents 
choosing one of the available destinations,} \rsa{our model is not standard with respect to the 
LQG MFG literature}. \jln{Specifically, our cost is non-convex and non-smooth in order to capture the 
combinatorial aspect of the discrete-decision making problem}.
\rsa{Hence, the existence proofs for a fixed point rely here on topological fixed point theorems rather than 
a contraction argument as in \cite{Huang07_Large}.} 
\jln{One of the main contributions of this paper is also to show that} \rsa{in the case of a uniform population, 
the infinite dimensional MFG fixed point problem \cite{huang2006large, Lasry07_MFG}
has a finite dimensional version that can be solved via Brouwer's fixed point theorem \cite{FunAnaCon}. 
For a nonuniform population, the existence of a fixed point path relies on an abstract fixed point theorem, 
namely Schauder's fixed point theorem \cite{FunAnaCon}. In both cases, to solve the MFG equation system, one needs 
to know the initial probability distribution of the players, whereas in the standard LQG MFG problems, 
it is sufficient to know the initial mean to anticipate the macroscopic behavior.} 
\rsa{Thus, in a nutshell, the theoretical tools needed to address this new formulation are thoroughly different.
Further highlighting the differences between the two problems, the standard LQ MFG problem with 
stochastic dynamics is entirely tractable, whereas solving an extension to the current formulation
with stochastic dynamics remains thus far beyond reach.
}

\jln{Preliminary versions of our results appeared in the conference papers \cite{Rab2014, Rab2015}.
Here we provide a unified discussion of our collective choice model for the deterministic and
stochastic scenarios, as well as more extensive results.
Many of the proofs were omitted from the conference papers due to space limitations 
and can be found here. The simulation section is also expanded with respect to \cite{Rab2014, Rab2015} 
and provides additional insight on the role of the different parameters in the model.
}
\subsection{Notation}

The following notation is used throughout the paper. We denote by $C(X,Y)$ the set of 
continuous functions from a normed vector space $X$ to $Y\subset \mathbb{R}^k$ with 
the standard supremum norm $\|.\|_\infty$. We fix a generic probability space 
$(\Omega,\mathscr{F},\mathbb{P})$ and denote by $\mathbb{P}(A)$ the probability 
of an event $A$, and by $\mathbb{E}(X)$ the expectation of a random variable $X$. 
The indicator function of a subset $X$ is denoted by $1_X$ and its
interior by $\overset{\circ}{X}$. We denote by $|X|$ the size of a finite set $X$. 
The transpose of a matrix $M$ is denoted by $M^T$. We denote by $I_k$ the identity 
$k\times k$ matrix. The subscript $i$ is used to index entities related to the agents, 
while the subscripts $j$ and $k$ are used to index entities related to the home destinations. 
We denote by $[x]_k$ the $k$-th component of a vector $x$.

\section{Tracking Problem and Basins of attraction} \label{section: tracking}

\subsection{Tracking Problem}
Following the MFG approach, we assume the trajectory $\bar{x}(t)$ in (\ref{eq:sysinde}) and (\ref{eq:sysin}) to be arbitrary for now and equal to $\hat{x}(t)$.
The cost functions (\ref{eq:sysinde}) and (\ref{eq:sysin}) can be written 
\begin{equation} \label{eq:track}
J_{i}(u_{i},\hat{x},x_i^0)=\min\limits_{j=1,\dots,l}  J_{ij}(u_{i},\hat{x},x_i^0) ,
\end{equation}
where
\iftoggle{jou}{
\begin{multline}\label{indivi_cost}
J_{ij}(u_{i},\hat{x},x_i^0)=    \int_0^T \; \left \{ \frac{q}{2} \| x_{i} - \hat{x} \|^{2} + 
\frac{r}{2} \| u_{i} \|^{2} \right \} \,\mathrm{d}t \\ 
+ \frac{M}{2} \| x_{i}(T)-p_{j} \|^{2} . 
\end{multline}}{
\begin{equation}\label{indivi_cost}
J_{ij}(u_{i},\hat{x},x_i^0)=    \int_0^T \; \left \{ \frac{q}{2} \| x_{i} - \hat{x} \|^{2} + 
\frac{r}{2} \| u_{i} \|^{2} \right \} \,\mathrm{d}t  
+ \frac{M}{2} \| x_{i}(T)-p_{j} \|^{2}  .
\end{equation}
}
Moreover, we have
\begin{equation*}
\inf\limits_{u_{i}(.)}  J_{i}(u_{i},\hat{x},x_i^0)
=\min\limits_{j=1,\dots,l} \left( \inf\limits_{u_{i}(.)} J_{ij}(u_{i},\hat{x},x_i^0)
 \right).
\end{equation*}
Assuming a full (local) state feedback, the optimal control for (\ref{eq:track}) is
\begin{align*}
u_{i}^{*} = u_{ij}^* && \text{if } J_{ij}(u_{ij}^{*},\hat{x},x_{i}^{0})= \min\limits_{k=1,\dots,l} J_{ik}(u_{ik}^*,\hat{x},x_{i}^{0}), 
\end{align*}
where $u_{ik}^*$ is the optimal solution of the simple linear quadratic tracking problem 
with cost function $J_{ik}$. 
We recall the optimal control laws \cite{anderson2007optimal}
\begin{align*}
u_{ik}^*(t) = - \frac{1}{r} B^T \Big(\Gamma(t) x_i + \beta_k(t)\Big ), && \forall k \in \{1,\dots,l\},
\end{align*}
with the corresponding optimal (simple) costs
\[
J_{ik}^{*}(\hat{x},x^{0}_i)=\frac{1}{2} (x_i^0)^T \Gamma (0)x_i^{0}+\beta_k(0)^{T} x_i^{0}+\delta_k (0),
\]
where $\Gamma$, $\beta_k$ and $\delta_k$ are respectively matrix-, vector-,
and real-valued functions satisfying the following backward propagating differential equations:
\begin{subequations}
\begin{align}
&\dot{\Gamma}-\frac{1}{r} \Gamma BB^{T}\Gamma +\Gamma A+A^{T}\Gamma +q I_{n}=0  \label{eq:abd-1} \\
&\dot{\beta_k}=\left(\frac{1}{r} \Gamma BB^{T}-A^{T}\right )\beta_k +q\hat{x} \label{eq:abd-2} \\
&\dot{\delta_k}=\frac{1}{2r} (\beta_k)^{T}BB^{T}\beta_k -\frac{1}{2}q  \hat{x}^{T}\hat{x},  \label{eq:abd-3} 
\end{align}
\end{subequations}
with the final conditions 
\[
\Gamma (T)=MI_{n}, \;\; \beta_k (T)=-Mp_k, \;\; \delta_k (T)=\frac{1}{2}M p_k^{T} p_k.
\]
\rsa{
We define the basins of attraction
\iftoggle{jou}{
\begin{multline} \label{eq:da}
D_j(\hat{x}) =  \Big \{x\in \mathbb R^n \text{ such that}\\  
 \big (\beta_{j}(0)-\beta_{k}(0) \big )^{T}x\leq \delta_{k}(0)-\delta_{j}(0), \, \forall k=1,\dots,l \Big \}, 
\end{multline}}{
\begin{equation} \label{eq:da}
D_j(\hat{x}) =  \Big \{x\in \mathbb R^n \text{ such that }  
 \big (\beta_{j}(0)-\beta_{k}(0) \big )^{T}x\leq \delta_{k}(0)-\delta_{j}(0), \, \forall k=1,\dots,l \Big \}, 
\end{equation}
}
for $j=1,\dots,l$.
If an agent $i$
is initially in $D_j(\hat{x})$, then the smallest 
optimal 
(simple) cost is 
$J_{ij}^*$, and player $i$ goes towards the corresponding destination 
point $p_j$.}
\begin{assumption} \label{assum-rule}
\rsa{Conventionally, we assume that if $x_i^0 \in \cap_{m=1}^{k} D_{j_m}(\hat{x})$, for some $j_1<\dots<j_k$, then the player $i$ goes towards $p_{j_1}$. Under Assumptions \ref{assumption: agent spread} and \ref{assumption: measure support}, this convention does not affect the analysis in case of random initial conditions.} 
\end{assumption}
We summarize the above analysis in the following lemma.

\begin{lem} \label{lemma: tracking}
Under Assumption \ref{assum-rule}, the tracking problem (\ref{eq:track}) has a unique optimal control law 
\begin{align}\label{eq:ctr}
u_{i}^{*}(t) = 
- \frac{1}{r} B^T \big (\Gamma(t) x_i + \beta_j(t) \big ) && \text{if } x^0_i \in D_j(\hat{x}), 
\end{align}
where $\Gamma$, $\beta_j$, $\delta_j$ are the unique solutions of (\ref{eq:abd-1})-(\ref{eq:abd-3}).
\end{lem} 

The optimal control laws (\ref{eq:ctr}) depend on the tracked path $\hat{x}(t)$  
and the local state $x_i$. 
As mentioned above, each agent should reach one of the predefined destinations. We show in the next lemma that for any horizon length T, M can be made large enough that each agent reaches an arbitrarily small neighborhood of some destination point by applying the control law (\ref{eq:ctr}).
\jln{The result is proved for tracked paths $\hat{x}(t)$ that 
are uniformly bounded with respect to $M$, a property that is shown to hold later in 
Lemma \ref{lemma:existance1} for the desired tracked paths (fixed point tracked paths).}
\begin{lem} \label{contr}
Suppose that the pair $(A,B)$ is controllable and for each $M>0$, the agents are optimally tracking 
a path $\hat{x}_M(t)$. \rsa{We suppose that the family $\hat{x}_M(t)$ is uniformly bounded with respect to $M$
for the norm $\Big(\int_0^T \|.\|^2\mathrm{dt}\Big)^{\frac{1}{2}}$.} Then, 
for any $\epsilon>0$, there exists $M_0>0$ such that for all $M>M_0$, 
each agent is at time $T$ in a ball of radius $\epsilon$ and centered 
at one of the $p_j$'s, for $j=1,\dots,l$.  
\end{lem}
\begin{IEEEproof}
See Appendix \ref{Proof1}.
\end{IEEEproof}
Given any continuous path $\hat{x}(t)$, there exist $l$ basins of attraction where all the agents initially in $D_j(\hat{x})$ prefer going towards $p_{j}$, $j=1,\dots,l$. Therefore, the mean of the population is highly dependent on the structure of $D_{j}(\hat{x})$, $j=1\dots,l$. In the next paragraph, we study the properties of these basins in more detail.

\subsection{Basins of Attraction} \label{section: separation}  

We start by giving an explicit solution of (\ref{eq:abd-2}) and (\ref{eq:abd-3}). Let $\Pi(t)=\frac{1}{r}\Gamma(t) BB^{T}-A^{T}$ and 
$\Phi(.,\eta)$, for $\eta \in \mathbb{R}$, be the unique solution of 
\begin{align} \label{TM}
\frac{d\Phi(t,\eta)}{dt}=\Pi(t)\Phi(t,\eta) && \Phi(\eta,\eta)=I_n, 
\end{align}
and 
\begin{equation} \label{exp-psi}
\Psi(\eta_1,\eta_2,\eta_3,\eta_4)=\Phi(\eta_1 , \eta_2)^TBB^{T}\Phi(\eta_3 , \eta_4).
\end{equation}
\rsa{Two main properties of the  
state transition matrix $\Phi$ are used in this paper, namely the matrix $\Phi(\eta_1 , \eta_2)$ has an inverse $\Phi(\eta_2 , \eta_1)$ and the  state transition matrix $\tilde{\Phi}(\eta_1,\eta_2)$
of $-\Pi^T$ is equal to $\Phi(\eta_2,\eta_1)^T$. For more details about the properties
of the state transition 
matrix, one can refer to 
\cite{Rugh1993}}. 
We have
\iftoggle{jou}{
\begin{equation} \label{eq:beta}
\begin{split}
\beta&_{k}(t)  =  -M\Phi(t,T)p_{k}+ q \int_T^t \;  \Phi(t,\sigma)\hat{x}(\sigma) \, \mathrm{d}\sigma  \\ 
\delta&_{k}(t)  =  \frac{1}{2}Mp_{k}^{T}p_{k} -\frac{q}{2}\int_T^t \;  \hat{x}(\sigma)^{T}\hat{x}(\sigma) \, \mathrm{d}\sigma   \\ 
& +\frac{M^{2}}{2r}p_{k}^{T}\int_T^t \;  \Psi(\eta,T,\eta,T) \, \mathrm{d}\eta \,p_{k}  \\  
&-\frac{Mq}{r}p_{k}^{T}\int_T^t \;  \int_T^\eta \;  \Psi(\eta,T,\eta,\sigma)\hat{x}(\sigma) \, \mathrm{d}\sigma \mathrm{d}\eta \\
&+ \frac{q^{2}}{2r}\int_T^t \; \int_T^\eta \; \int_T^\eta \;
\hat{x}(\sigma)^{T}\Psi(\eta,\sigma,\eta,\tau)\hat{x}(\tau) 
\,  \mathrm{d}\tau \mathrm{d}\sigma \mathrm{d}\eta. 
\end{split}
\end{equation}}{
\begin{equation} \label{eq:beta}
\begin{split}
\beta&_{k}(t)  =  -M\Phi(t,T)p_{k}+ q \int_T^t \;  \Phi(t,\sigma)\hat{x}(\sigma) \, \mathrm{d}\sigma  \\ 
\delta&_{k}(t)  =  \frac{1}{2}Mp_{k}^{T}p_{k} -\frac{q}{2}\int_T^t \;  \hat{x}(\sigma)^{T}\hat{x}(\sigma) \, \mathrm{d}\sigma   \\ 
& +\frac{M^{2}}{2r}p_{k}^{T}\int_T^t \;  \Psi(\eta,T,\eta,T) \, \mathrm{d}\eta \,p_{k}   
-\frac{Mq}{r}p_{k}^{T}\int_T^t \;  \int_T^\eta \;  \Psi(\eta,T,\eta,\sigma)\hat{x}(\sigma) \, \mathrm{d}\sigma \mathrm{d}\eta \\
&+ \frac{q^{2}}{2r}\int_T^t \; \int_T^\eta \; \int_T^\eta \;
\hat{x}(\sigma)^{T}\Psi(\eta,\sigma,\eta,\tau)\hat{x}(\tau) 
\,  \mathrm{d}\tau \mathrm{d}\sigma \mathrm{d}\eta. 
\end{split}
\end{equation}
}
By replacing (\ref{eq:beta}) in the expression of $D_j(\hat{x})$, (\ref{eq:da}) can be written 
\iftoggle{jou}{
\begin{multline} \label{eq:dae}
D_j(\hat{x}) = \Big \{ x\in \mathbb R^n \text{ such that}\\ \beta_{jk}^{T}x \leq \delta_{jk} + \alpha_{jk}(\hat{x}),\, \forall k=1,\dots,l \Big \}, 
\end{multline}}{
\begin{equation} \label{eq:dae}
D_j(\hat{x}) = \Big \{ x\in \mathbb R^n \text{ such that } \beta_{jk}^{T}x \leq \delta_{jk} + \alpha_{jk}(\hat{x}),\, \forall k=1,\dots,l \Big \}, 
\end{equation}
}
where
\iftoggle{jou}{
\begin{equation} \label{eq:de0}
\begin{split}
\beta_{jk} & = M\Phi(0,T)(p_{k}-p_{j})\\
\delta_{jk} & = \frac{1}{2}Mp_{k}^{T}p_{k}-\frac{1}{2}Mp_{j}^{T}p_{j} \\
&+\frac{M^{2}}{2r}p_{k}^{T} \int_T^0 \;  \Psi(\eta,T,\eta,T) \, \mathrm{d}\eta \,p_{k}\\ 
&-\frac{M^{2}}{2r}p_{j}^{T} \int_T^0 \;  \Psi(\eta,T,\eta,T) \, \mathrm{d}\eta \,p_{j} \\
\alpha_{jk}(\hat{x})&  =
\frac{Mq}{r}(p_{j}-p_{k})^{T}
\int_T^0 \! \int_T^\eta \;  
\Psi(\eta,T,\eta,\sigma)\hat{x}(\sigma) \, \mathrm{d}\sigma \mathrm{d}\eta. 
\end{split} 
\end{equation}}{
\begin{equation} \label{eq:de0}
\begin{split}
\beta_{jk} & = M\Phi(0,T)(p_{k}-p_{j})\\
\delta_{jk} & = \frac{1}{2}Mp_{k}^{T}p_{k}-\frac{1}{2}Mp_{j}^{T}p_{j} \\
&+\frac{M^{2}}{2r}p_{k}^{T} \int_T^0 \;  \Psi(\eta,T,\eta,T) \, \mathrm{d}\eta \,p_{k}
-\frac{M^{2}}{2r}p_{j}^{T} \int_T^0 \;  \Psi(\eta,T,\eta,T) \, \mathrm{d}\eta \,p_{j} \\
\alpha_{jk}(\hat{x})&  =
\frac{Mq}{r}(p_{j}-p_{k})^{T}
\int_T^0 \! \int_T^\eta \;  
\Psi(\eta,T,\eta,\sigma)\hat{x}(\sigma) \, \mathrm{d}\sigma \mathrm{d}\eta. 
\end{split} 
\end{equation}
}

\section{Fixed Point - Deterministic Case} \label{section: fixed}

Having presented the solution of the general tracking problem, we now seek 
a continuous path $\hat{x}(t)$ that is sustainable, in the sense that it can 
be replicated by the mean of the agents under their optimal tracking 
control laws. We start by analyzing the finite size population where the 
initial state of each agent is \jln{known to all the agents}. 
We start our search for the desired path $\hat{x}(t)$ by computing the mean $\bar{x}(t)$ when 
tracking any continuous path $\hat{x}(t)$. The dynamics of the mean when 
tracking $\hat{x}\in C([0,T],\mathbb{R}^{n})$ satisfies
\iftoggle{jou}{
\begin{multline} \label{eq:muapp1}
\dot{\bar{x}} = -\Pi^{T}\bar{x} -\frac{q}{r}BB^{T}\int_T^t \; \ \Phi(t,\sigma)\hat{x}(\sigma)\, \mathrm{d}\sigma \\
+\frac{M}{r}BB^{T}\Phi(t,T)p_{\lambda(\hat{x})},   \end{multline}}
{
\begin{equation} \label{eq:muapp1}
\dot{\bar{x}} = -\Pi^{T}\bar{x} -\frac{q}{r}BB^{T}\int_T^t \; \ \Phi(t,\sigma)\hat{x}(\sigma)\, \mathrm{d}\sigma
+\frac{M}{r}BB^{T}\Phi(t,T)p_{\lambda(\hat{x})}  , \end{equation}
}
where $\bar{x}(0)=\frac{1}{N}\sum_{i=1}^{N}x_i^0\triangleq\bar{x}_{0}$,  $p_{\lambda(\hat{x})}=\sum_{j=1}^l\frac{\lambda_j(\hat{x})}{N} p_{j}$ and $\lambda_j(\hat{x})$ is the number of agents initially in $D_j(\hat{x})$, which therefore pick $p_j$ as a destination. We obtain (\ref{eq:muapp1}) by substituting (\ref{eq:beta}) in (\ref{eq:ctr}) and the resulting control law in (\ref{eq: agents dynamics}) to subsequently compute $\bar{x} = 1/N\sum_{i=1}^N x_i$ and its derivative. Thus, the mean of the population $\bar{x}$ when tracking any continuous path $\hat{x}$ is the image of $\hat{x}$ by a composite map $G=G_2\circ G_1$, where
\begin{align*}
G_1: C([0,T],\mathbb{R}^n) &\mapsto C([0,T],\mathbb{R}^n)\times\mathbb{N}^l \\
\hat{x} &\mapsto \Big (\hat{x},\big(\lambda_1(\hat{x}),\dots,\lambda_l(\hat{x})\big) \Big ) \\
G_2: C([0,T],\mathbb{R}^n)\times\mathbb{N}^l  & \mapsto C([0,T],\mathbb{R}^n) \\
\Big (\hat{x},\big(\lambda_1,\dots,\lambda_l\big) \Big ) &\mapsto \bar{x}
\end{align*}
and $\bar{x}=G_2\bigg(\Big (\hat{x},\big(\lambda_1,\dots,\lambda_l\big) \Big ) \bigg)$ is the unique solution of (\ref{eq:muapp1}) in which $\lambda_j(\hat{x})$ is equal to an arbitrary $\lambda_j$, $j=1,\dots,l$. 

The desired path \jln{describing the mean trajectory} is a fixed point of $G$. 
\rsa{In the following, we construct a one to one map between the fixed points of $G$
and the fixed points of a \emph{finite dimensional} operator $F$ describing the way 
the population splits between the destination points. We start by showing that the fixed
points of $G$ have a special form.} 
For any $\lambda=(\lambda_1,\dots,\lambda_l)\in \{0,...,N\}^l$, we define a new map $T_{\lambda}$ from $C([0,T],\mathbb{R}^n)$ to $C([0,T],\mathbb{R}^n)$, where $T_{\lambda}(\hat{x})=G_2(\hat{x},\lambda)$.
\rsa{If $\hat{x}$ is a fixed point of
$G$ and $\lambda_0=\lambda(\hat{x})$, then $\hat{x}$ is a fixed point of 
$T_{\lambda_0}$. In the following lemma, we show that for any $\lambda$, $T_\lambda$ has a unique 
fixed point $y_\lambda$, and we give an explicit form for $y_\lambda$.}  
\begin{lem} \label{lemma:existance1}
For all $\lambda=(\lambda_1,\dots,\lambda_l) \in \{0,...,N\}^l$, 
$T_{\lambda}$ has a unique fixed point equal to
\begin{align} \label{eq:fixpoint}
y_\lambda=R_{1}(t)\bar{x}_{0}+R_{2}(t)p_{\lambda},
\end{align}
where
\rsa{\begin{equation} \label{eq:r1r2}
\begin{split}
R_1(t)&=\Phi_P(t,0)\\
R_2(t)&=\frac{M}{r}\int_0^t 
\Phi_{P} (t,\sigma)BB^T
\Phi_{P} (T,\sigma)^T\mathrm{d}
\sigma ,
\end{split}
\end{equation}
and $P$ and $\Phi_{P} (t,\eta)$ are the unique solutions of
\begin{equation} \label{aux_riccati}
\begin{split}
&\dot{P}=-PA-A^TP+\frac{1}{r}PBB^TP, \quad P(T)=MI_n\\
&\dot{\Phi}_{P} (t,\eta) = -(A-\frac{1}{r}BB^TP)^T\Phi_{P}(t,\eta), \quad \Phi_{P}(\eta,\eta)=I_n.
\end{split}
\end{equation}
Moreover, if $(A,B)$ is controllable, then the paths $y_\lambda$ are uniformly 
bounded with respect to $M$ for the norm $\Big(\int_0^T \|.\|^2\mathrm{dt}\Big)^{\frac{1}{2}}$.}
\end{lem}
\begin{IEEEproof}
See Appendix \ref{Proof1}.
\end{IEEEproof} 
\rsa{The fixed point path (\ref{eq:fixpoint}) is the 
optimal state of the LQR problem (\ref{indivi_cost}), where $q=0$ and the final destination point is $p_\lambda$.}

\rsa{Hitherto, we know that the fixed points of $G$ are of the form (\ref{eq:fixpoint}). To 
narrow the search further, we derive
a necessary condition on the vector
$\lambda$ so that the corresponding
$y_\lambda$ is a fixed point of $G$. 
We start by replacing the new expression of the fixed points  (\ref{eq:fixpoint}) in the expressions 
of the basins of attraction, which then have the following form:
}
\iftoggle{jou}{
\begin{multline*}
H_j^\lambda=D_j(R_{1}(t)\bar{x}_{0}+R_{2}(t)p_{\lambda})=\\
\{x\in \mathbb{R}^n| \beta_{jk}^{T}x \leq \delta_{jk} + \theta_{jk}\bar{x}_0 + \xi_{jk} p_{\lambda} \,\, \forall k=1,\dots,l \big \},
\end{multline*}}{
\begin{equation*}
H_j^\lambda=D_j(R_{1}(t)\bar{x}_{0}+R_{2}(t)p_{\lambda})=
\{x\in \mathbb{R}^n|\beta_{jk}^{T}x \leq \delta_{jk} + \theta_{jk}\bar{x}_0 + \xi_{jk} p_{\lambda} \,\, \forall k=1,\dots,l \big \},
\end{equation*}
}
where
\begin{align} \label{theta1}
\theta_{jk}&=\frac{Mq}{r}(p_{j}^{T}-p_{k}^{T})
\int_T^0 \!  \ \int_T^\eta \! \Psi(\eta,T,\eta,\sigma)R_{1}(\sigma)
 \, \mathrm{d}\sigma \mathrm{d}\eta 
\end{align}
\begin{align} \label{theta2}
\xi_{jk}&=\frac{Mq}{r}(p_{j}^{T}-p_{k}^{T})
\int_T^0 \!  \ \int_T^\eta \! \Psi(\eta,T,\eta,\sigma)R_{2}(\sigma)
 \, \mathrm{d}\sigma \mathrm{d}\eta.
\end{align}
\rsa{Following the discussion above and 
Lemma \ref{lemma:existance1}, we can 
claim that if $\hat{x}$ is a fixed point
of $G$, then $\hat{x}$ is of the form (\ref{eq:fixpoint}), where 
\begin{align} 
\lambda &= \lambda(\hat{x})=(\big | \big\{ x_i^0| x_i^0 \in D_1(\hat{x})\}
\big |,\dots,\big | \big\{ x_i^0| x_i^0 \in D_l(\hat{x})\}\big |) \nonumber \\
&= (\big | \big\{ x_i^0| x_i^0 \in H_1^\lambda\}
\big |,\dots,\big | \big\{ x_i^0| x_i^0 \in H_l^\lambda\}\big |)\triangleq F(\lambda).
\label{finiteop}
\end{align}
Thus, we proved that if $\hat{x}$ is a fixed point of $G$, then $\hat{x}$ is of the
form (\ref{eq:fixpoint}), where $\lambda$ is a fixed point of the finite dimensional
operator $F$ defined in (\ref{finiteop}). 
To prove the converse, we consider a fixed point $\lambda$ of $F$ 
and the path $\hat{x}=R_{1}(t)\bar{x}_{0}+R_{2}(t)p_{\lambda}$. We have
\begin{align*}
\lambda&=F(\lambda)
=(\big | \big\{ x_i^0| x_i^0 \in H_1^\lambda\}
\big |,\dots,\big | \big\{ x_i^0| x_i^0 \in H_l^\lambda\}\big |)\\
&=(\big | \big\{ x_i^0| x_i^0 \in D_1(\hat{x})\}
\big |,\dots,\big | \big\{ x_i^0| x_i^0 \in D_l(\hat{x})\}\big |)=\lambda(\hat{x}),
\end{align*}
where the third equality is a consequence of the form of $\hat{x}$.
The path $\hat{x}$ is the unique fixed point of 
$T_\lambda$. But $\hat{x}=T_\lambda(\hat{x})=T_{\lambda(\hat{x})}(\hat{x})=G(\hat{x})$. 
Therefore, $\hat{x}$ is a fixed point of $G$.}
\rsa{We summarize the above discussion in the following theorem.
\begin{thm} \label{theorem:fixedl}
The path $\hat{x}$ is a fixed point of $G$ if and only if it has the form (\ref{eq:fixpoint}), 
where $\lambda$ is a fixed point of $F$.
\end{thm}  }

Without loss of generality, we can index
in the binary choice case $(l=2)$ 
the agents going 
towards $p_1$ by numbers lower than those given to the agents going towards $p_2$ as follows:
\begin{align} \label{eq:index}
\beta_{12}^{T}x_{1}^{0}\leq \beta_{12}^{T}x_{2}^{0}\leq \dots \leq \beta_{12}^{T}x_{N}^{0}.
\end{align}
Then, the necessary and sufficient condition for the existence of the desired path reduces to 
a simple inequality as shown in the following theorem.
\begin{thm} \label{theorem:fixed1}
For $l=2$, the following statements hold:
\begin{enumerate}
\item $\hat{x}$ is a fixed point of $G$ if and only if there exists a seperating $\alpha$ in $\{0,...,N\}$ such that:
\iftoggle{jou}{
\begin{multline} \label{eq:ineq}
(\beta_{12})^{T}x_{\alpha}^{0}-\delta_{12}-\theta_{12}\bar{x}_0-\xi_{12} p_2\leq 
\frac{\alpha}{N} \xi_{12}(p_1-p_2)\\ < (\beta_{12})^{T}x_{\alpha +1}^{0}-\delta_{12}-\theta_{12}\bar{x}_0-\xi_{12}p_2. 
\end{multline}}
{
\begin{flalign} \label{eq:ineq}
&\text{For $\alpha$ different from $0$ and $N$, } & \nonumber \\
&\beta_{12}^{T}x_{\alpha}^{0}-\delta_{12}-\theta_{12}\bar{x}_0-\xi_{12}p_2\leq 
\frac{\alpha}{N} \xi_{12}(p_1-p_2) < \beta_{12}^{T}x_{\alpha +1}^{0}-\delta_{12}-\theta_{12}\bar{x}_0-\xi_{12}p_2, &
\end{flalign}
}
\begin{flalign} \label{eq:ineq1}
\text{For $\alpha=0$, } \quad & 0 < (\beta_{12})^{T}x_{ 1}^{0}-\delta_{12}-\theta_{12}\bar{x}_0-\xi_{12}p_2, &
\end{flalign}
\begin{flalign} \label{eq:ineq2}
\text{For $\alpha=N$, } \quad & (\beta_{12})^{T}x_{N}^{0}-\delta_{12}-\theta_{12}\bar{x}_0-\xi_{12}p_2\leq 0. &
\end{flalign}
In this case, $\alpha$ is the number of agents that go towards $p_1$.
\item For $\xi_{12}(p_1-p_2) \geq 0$, there exists $\alpha$ in $\{0,...,N\}$ satisfying (\ref{eq:ineq}), 
(\ref{eq:ineq1}) or (\ref{eq:ineq2}).
\item For $\xi_{12}(p_1-p_2) <0 $, there exists at most one $\alpha$ in $\{0,...,N\}$ satisfying (\ref{eq:ineq}), 
(\ref{eq:ineq1}) or (\ref{eq:ineq2}). Moreover, there exist some initial distributions 
for which no such $\alpha$ exists.
\end{enumerate}
\end{thm}
\begin{IEEEproof}
See Appendix \ref{Proof1}.
\end{IEEEproof} 
\begin{rem}
For the scalar case ($n=1$), $\xi_{12}(p_1-p_2)$ is always non-negative. 
In fact, in this case, $\Phi$ and $\Phi_P$
are real exponential functions. This implies that $\xi_{12}(p_1-p_2) \geq 0$.
\end{rem}

Theorem \ref{theorem:fixedl} shows that computing the anticipated macroscopic behaviors (fixed points of $G$) is equivalent to computing all the
fixed points $\lambda$'s of $F$ for which it is necessary to assume that each agent knows the exact initial states of all the other agents. Thus, for each $\lambda \in \{0,\dots,N\}^l$ such that $1^T\lambda/N=1$, the agents must count the number of initial positions $\eta_j$ inside each region $H_j^\lambda$, $j=1,\dots,l$. If $\eta_j=\lambda_j$, for $j=1,\dots,l$, then $\lambda$ is a fixed point
of $F$. The map $F$ may have multiple fixed points. Hence, an a priori agreement on how to choose $\lambda$ should exist. For example, although non-cooperative, the agents may anticipate that their majority will look for the most socially favorable Nash equilibrium if many exist and $N$ is large. This $\lambda$ corresponds to  minimizing the total cost $\frac{1}{N}\sum_{i=1}^NJ_i\Big(u_i^*(x_i,\hat{x}),\hat{x}),x_i^0\Big)=\frac{1}{N}\sum_{i=1}^N\min\limits_{k=1,\dots,l}\Big \{\frac{1}{2} (x_i^0)^T \Gamma (0)x_i^{0}+\beta_k(0)^{T} x_i^{0}+\delta_k (0)\Big \}$, which is also computable by just knowing the exact initial conditions of all the agents. Once the agents agree on a $\lambda$, they start tracking the corresponding fixed point defined by (\ref{eq:fixpoint}).
The fixed point vector $\lambda$ describes the
way the population splits between the destination points. In fact, $\lambda_j$,
$j=1,\dots,l$, is 
the number of agents that go towards $p_j$.
When $N$ is large, this algorithm is costly in terms of number of counting and verification operations. In the next section, we consider the limiting case of a large population with random initial conditions. 

\section{Fixed Point - Stochastic Case} \label{section: fixedpointstoch}

In this section, we assume that the agents' initial conditions $x^0_i$
are random and i.i.d. on some probability space $(\Omega,\mathscr{F},\mathbb{P})$ 
with distribution $P_0$ on $\mathbb R^n$. In this case, we show that 
for a large population, it is enough to know $P_0$ to anticipate the 
macroscopic behavior.
For a continuous path $\hat{x}$ and for all $\omega \in \Omega$, 
we denote by $\Lambda_{jN}(\omega)$ the number of $x_i^0(\omega)$ in $D_j(\hat{x})$, 
by $\bar{x}_N(\omega)$ the mean of the population when tracking $\hat{x}$, 
and by $\bar{x}_\infty$ the limit with probability one of $\bar{x}_N$ 
as $N$ goes to infinity. 
We deduce from (\ref{eq:muapp1}) that for all $\omega$ in $\Omega$,
\iftoggle{jou}{
\begin{multline*}
\bar{x}_N(\omega)  =\Phi(0,t)^T\frac{\sum_{i=1}^{N}x^0_i(\omega)}{N}\\
+\sum_{j=1}^l\frac{\Lambda_{jN}(\omega)}{N}\frac{M}{r}\int_0^t \! \ 
\Psi(\sigma,t,\sigma,T)p_{j} \, \mathrm{d}\sigma \\
 -\frac{q}{r}\int_0^t \! \ \int_T^\sigma \! \ \Psi(\sigma,t,\sigma,\tau)\hat{x}(\tau) \, \mathrm{d}\tau \mathrm{d}\sigma.
\end{multline*}}{
\begin{multline*}
\bar{x}_N(\omega)  =\Phi(0,t)^T\frac{\sum_{i=1}^{N}x^0_i(\omega)}{N}
+\sum_{j=1}^l\frac{\Lambda_{jN}(\omega)}{N}\frac{M}{r}\int_0^t \! \ \Psi(\sigma,t,\sigma,T)p_{j} \, \mathrm{d}\sigma\\
 -\frac{q}{r}\int_0^t \! \ \int_T^\sigma \! \ \Psi(\sigma,t,\sigma,\tau)\hat{x}(\tau) \, \mathrm{d}\tau \mathrm{d}\sigma,
\end{multline*}
}
where $\Psi$ is defined in (\ref{exp-psi}).
By the strong Law of large numbers, 
$\frac{\Lambda_{jN}}{N}=\frac{1}{N}\sum_{i=1}^{N}1_{D_j(\hat{x})}(x_i^0)$ 
and $\frac{\sum_{i=1}^{N}x^0_i}{N}$ converge
with probability one respectively to 
$P_0(D_j(\hat{x})) = \mathbb{P}\Big (x_i^0 \in D_j (\hat{x})\Big)$ 
and $\mathbb{E}x^0_i\triangleq\mu_0$, as $N$ goes to infinity.
Hence, 
\iftoggle{jou}{
\begin{multline} \label{eq:stdt}
 \bar{x}_\infty  =\Phi(0,t)^T\mu_0\\
+\sum_{j=1}^lP_0(D_j(\hat{x}))\frac{M}{r}\int_0^t \! \ \Psi(\sigma,t,\sigma,T)p_{j} \, \mathrm{d}\sigma  \\ 
 -\frac{q}{r}\int_0^t \! \ \int_T^\sigma \! \ \Psi(\sigma,t,\sigma,\tau)\hat{x}(\tau) \, \mathrm{d}\tau \mathrm{d}\sigma. 
\end{multline}}{
\begin{multline} \label{eq:stdt}
 \bar{x}_\infty  =\Phi(0,t)^T\mu_0
+\sum_{j=1}^lP_0(D_j(\hat{x}))\frac{M}{r}\int_0^t \! \ \Psi(\sigma,t,\sigma,T)p_{j} \, \mathrm{d}\sigma \\
 -\frac{q}{r}\int_0^t \! \ \int_T^\sigma \! \ \Psi(\sigma,t,\sigma,\tau)\hat{x}(\tau) \, \mathrm{d}\tau \mathrm{d}\sigma. 
\end{multline}
}
%
%
Equation (\ref{eq:stdt}) defines an operator $G_s$ that maps the tracked path 
$\hat{x}$ to the mean $\bar{x}_\infty$. This operator and its fixed points, if any, 
depend only on the initial statistical distribution of the agents. 
%
\jln{The limiting equation (\ref{eq:stdt}) also corresponds to the following stochastic problem.}
Assume that the only public information is the initial statistical distribution. 
As in the deterministic case, we start our search for a fixed point path by replacing 
$\bar{x}$ in (\ref{eq:sysin}) by a continuous path $\hat{x}$. By Lemma \ref{lemma: tracking}, 
there exist $l$ regions $D_{j}(\hat{x})$ such that the agents initially in $D_{j}(\hat{x})$ 
select the control law (\ref{eq:ctr})
when tracking $\hat{x}$. By substituting (\ref{eq:beta}) in (\ref{eq:ctr}) and 
the resulting control law in (\ref{eq: agents dynamics}), we show that the mean 
trajectory $\mathbb{E}(x_i)$ of a generic agent is equal to $G_s(\hat{x})$. 

The next theorem establishes the existence of a fixed point of $G_s$. 
We define the set $\Delta_l=\{(\lambda_1,\dots,\lambda_l)\in [0,1]^l| \sum_{j=1}^l\lambda_j=1\}$ and the map $F_s$ from $\Delta_l$ into itself such that 
\iftoggle{jou}{
\begin{multline*}
F_s(\lambda_1,\dots,\lambda_l)=\\
 \begin{bmatrix}
\mathbb{P}\Big (\beta_{1j}^T x_i^0 \leq \delta_{1j} + \theta_{1j}\mu_{0} + \xi_{1j}p_\lambda,\, \forall j=1,\dots,l\Big)\\
\vdots\\
\mathbb{P}\Big (\beta_{lj}^T x_i^0 \leq \delta_{lj} + \theta_{lj}\mu_{0} + \xi_{lj}p_\lambda,\, \forall j=1,\dots,l\Big) 
\end{bmatrix}^T,
\end{multline*}}{
\begin{equation*}
F_s(\lambda_1,\dots,\lambda_l)=
 \begin{bmatrix}
\mathbb{P}\Big (\beta_{1j}^Tx_i^0 \leq \delta_{1j} + \theta_{1j}\mu_{0} + \xi_{1j}p_\lambda,\, \forall j=1,\dots,l\Big)\\
\vdots\\
\mathbb{P}\Big (\beta_{lj}^Tx_i^0 \leq \delta_{lj} + \theta_{lj}\mu_{0} + \xi_{lj}p_\lambda,\, \forall j=1,\dots,l\Big) 
\end{bmatrix}^T,
\end{equation*}
}
where $p_\lambda=\sum_{k=1}^l\lambda_k p_k$. 
The quantities $\beta_{kj}$ and $\delta_{kj}$ are defined in (\ref{eq:de0}), 
and $\theta_{kj}$ and $\xi_{kj}$ are defined in (\ref{theta1}) and (\ref{theta2}). 
%
\begin{assumption}	\label{assumption: agent spread}
We assume that $P_0$ is such that the $P_0$-measure of hyperplanes is zero. 
\end{assumption}
\begin{thm} \label{lemma:existance11}
Under Assumption \ref{assumption: agent spread}, the following statements hold: 
\begin{enumerate}[(i)]
\item $\hat{x}$ is a fixed point of $G_s$ if and only if there exists $\lambda=(\lambda_1,\dots,\lambda_l)$ in $\Delta_l$ such that 
\begin{align}\label{stineq-multi}
F_s(\lambda)=\lambda,
\end{align} 
for $\hat{x}(t)=R_1(t)\mu_0+R_2(t)p_\lambda$. 
\item $F_s$ has a\jln{t least one} fixed point (equivalently $G_s$ has a\jln{t least one} fixed point).
\item For $l=2$, if $\xi_{12}(p_1-p_2) \leq 0 $, then $G_s$ has a unique fixed point.
\end{enumerate}
\end{thm}
\begin{IEEEproof}
See Appendix \ref{Proof2}.
\end{IEEEproof}
\rsa{The finite dimensional operators $F$ and $F_s$
defined respectively in the deterministic and
stochastic cases have similar structures. In fact,
in the deterministic case, if 
the sequence $\{x_i^0\}_{i=1}^N$ of initial conditions
is interpreted as a random variable on some 
probability space $(\Omega,\mathcal{F},\mathbb{P})$
with distribution $P_0(\mathcal{A})=1/N\sum_{i=1}^N 1_{\{x_i^0 \in \mathcal{A}\}}$, for all (Borel) 
measurable sets $\mathcal{A}$, then 
$F(\lambda)=NF_s(\lambda/N)$.}
 
In Theorem  \ref{lemma:existance11}, (i) shows that computing the anticipated macroscopic behaviors is equivalent to computing all the vectors $\lambda$ satisfying (\ref{stineq-multi}) under the corresponding constraint on $\hat{x}$. To compute a $\lambda$ satisfying (\ref{stineq-multi}), each agent is assumed to know the initial statistical distribution of the agents. As in the deterministic case, multiple $\lambda$'s may exist. Hence, an a priori agreement on how to choose $\lambda$ should exist. In that respect, the agents may implicitly assume that collectively they will opt for the $\lambda$ (assuming it is unique!) that minimizes the total expected population cost  
\iftoggle{jou}{
\begin{multline*}
\mathbb{E}J_i\Big(u_i^*(x_i,\hat{x}),\hat{x}),x_i^0\Big)=\\
\mathbb{E}\min\limits_{k=1,\dots,l}\Big \{\frac{1}{2} (x_i^0)^T \Gamma (0)x_i^{0}+\beta_k(0)^{T} x_i^{0}+\delta_k (0)\Big \}.
\end{multline*} }
{
\begin{equation*}
\mathbb{E}J_i\Big(u_i^*(x_i,\hat{x}),\hat{x}),x_i^0\Big)=
\mathbb{E}\min\limits_{k=1,\dots,l}\Big \{\frac{1}{2} (x_i^0)^T \Gamma (0)x_i^{0}+\beta_k(0)^{T} x_i^{0}+\delta_k (0)\Big \},
\end{equation*}
}
which can be evaluated if the agents know the initial statistical distribution of the population.

\subsection{Computation of The Fixed Points} \label{section: promoters}

The map $F_s$ is not necessarily a contraction. Hence, it is sometimes 
impossible to compute its fixed points by the simple iterative method 
$\lambda_{k+1}=F_s(\lambda_k)$.

\subsubsection{Binary Choice Case}

We give two simple methods to compute a fixed point of $F_s$ in the binary choice case. The first method is applicable if $\xi_{12}(p_1-p_2) > 0 $. We define in $[0,1]$ a sequence $\alpha_k$ such that $\alpha_0$ is an arbitrary number in $[0,1]$ and
\begin{equation} \label{tititi}
\lambda_{k+1}=(\alpha_{k+1},1-\alpha_{k+1})=F_s(\alpha_{k},1-\alpha_{k})=F_s(\lambda_k).
\end{equation}
Given that $\xi_{12}(p_1-p_2) > 0 $,  $\Big[F_s(t,1-t) \Big]_1$ increases with $t$. We show by induction that $\alpha_k$ is monotone. But $\alpha_k \in [0,1]$, therefore, $\alpha_k$ converges to some limit $\alpha$. By the continuity of $F_s$, $(\alpha,1-\alpha)$ satisfies (\ref{stineq-multi}). 
Since in this case $F_s$ may have multiple fixed points,  the $\lambda=(\alpha,1-\alpha)$ 
obtained using this approach depends on the initial value $\lambda_0=(\alpha_0,1-\alpha_0)$. 
If we define 
\begin{equation} \label{teeee}
\bar{x}_k=R_{1}(t)\mu_{0}+R_{2}(t)p_{\lambda_k}. 
\end{equation}
This sequence converges to a fixed point of $G_s$. The second method is applicable if $\xi_{12}(p_1-p_2) \leq 0 $. In this case $\Big[F_s(\lambda,1-\lambda) \big]_1-\lambda$ decreases with $\lambda$. Hence, one can compute
the unique zero of this function by the bisection method.

\subsubsection{General Case}

In general ($l>2$), $F_s$ is a vector of probabilities of some regions 
delimited by hyperplanes. 
Although a fixed point could be computed using Newton's method, this is
computationally expensive as it requires the values of the inverse 
of the Jacobian matrix at the root estimates. Alternatively, one can compute 
a fixed point of $F_s$ using a quasi Newton method such as
Broyden's method \cite{broyden1965class} ( see Section \ref{section: simulation}). 
Using this method, the  inverse of the Jacobian  can be estimated 
recursively provided that $F_s$ is continuously differentiable; 
this will be the case if the initial probability distribution has a continuous 
probability density function.

\subsection{Gaussian Binary Choice Case}	\label{Gaussian binary}

We showed in Theorem \ref{lemma:existance11} that for the binary choice case ($l=2$), 
if $\xi_{12}(p_1-p_2)< 0$, then $G_s$ has a unique fixed point. 
We now prove that for the binary choice case and Gaussian initial distribution irrespective of the sign of $\xi_{12}(p_1-p_2)$, $G_s$ has a unique fixed point 
provided that the initial spread of the agents is ``sufficient". For any $n\times n$ matrix $\Sigma_0$ such that $(\beta_{12})^T\Sigma_0\beta_{12} < \big(\xi_{12}(p_1-p_2)\big)^2/2\pi$, we define
\iftoggle{jou}{
\begin{multline*}
a(\Sigma_0)=\delta_{12}+\xi_{12}p_2-\\
\sqrt{2(\beta_{12})^T\Sigma_0\beta_{12}}\sqrt{\log \xi_{12}(p_1-p_2) -\frac{1}{2}\log2\pi(\beta_{12})^T\Sigma_0\beta_{12} }
\end{multline*}
\begin{multline*}
b(\Sigma_0)=\delta_{12}+\xi_{12}p_1+\\
\sqrt{2(\beta_{12})^T\Sigma_0\beta_{12}}\sqrt{\log \xi_{12}(p_1-p_2) -\frac{1}{2}\log2\pi(\beta_{12})^T\Sigma_0\beta_{12} }
\end{multline*}}{
\begin{equation*}
a(\Sigma_0)=\delta_{12}+\xi_{12}p_2-
\sqrt{2(\beta_{12})^T\Sigma_0\beta_{12}}\sqrt{\log \xi_{12}(p_1-p_2) -\frac{1}{2}\log2\pi(\beta_{12})^T\Sigma_0\beta_{12} }
\end{equation*}
\begin{equation*}
b(\Sigma_0)=\delta_{12}+\xi_{12}p_1+
\sqrt{2(\beta_{12})^T\Sigma_0\beta_{12}}\sqrt{\log \xi_{12}(p_1-p_2) -\frac{1}{2}\log2\pi(\beta_{12})^T\Sigma_0\beta_{12} }
\end{equation*}
}
\[
S(\Sigma_0)=\Big\{\mu_0 \in \mathbb{R}^n,\, \big (\beta_{12}^T-\theta_{12}\big)\mu_0 \in (a(\Sigma_0),b(\Sigma_0))\Big\}.
\]
\begin{thm} \label{Flat} 
$G_s$ has a unique fixed point if at least one of the following conditions is satisfied:
\begin{enumerate}
\item $\beta_{12}^T\Sigma_0\beta_{12}\geq \frac{\big(\xi_{12}(p_1-p_2)\big)^2}{2\pi}$.
\item $\mu_0 \notin S(\Sigma_0)$.
\end{enumerate}
\end{thm}
\begin{IEEEproof}
See Appendix \ref{Proof2}.
\end{IEEEproof}

Theorem \ref{Flat} states that in the Gaussian binary choice case, if the initial 
distribution of the agents has enough spread,
then the agents will anticipate the collective 
behavior in a unique way. On the other hand, if the uncertainty in their initial positions 
is low enough and the mean of population is inside the region $S(\Sigma_0)$ (a region 
delimited by two parallel hyperplanes), then the agents can anticipate the collective behavior 
in multiple ways. 

\section{Nonuniform Population with Initial Preferences} \label{section: initialpreferences}

Hitherto, the agents' initial affinities towards different potential targets 
are dictated only by their initial positions in space. In this section, 
the model is further generalized by considering that in addition to their 
initial positions, the agents are affected by their a priori opinion. 
When modeling smoking decision in schools for example \cite{nakajima2007measuring},
this could represent a teenager's tendency towards ``Smoking" or ``Not Smoking", which 
is the result of some endogenous factors such as parental pressure, financial condition, 
health, etc. When modeling elections, this would reflect personal  preferences that 
transcend party lines. Moreover, we assume in this section that the agents 
have nonuniform dynamics. 

We consider $N$ agents with nonuniform dynamics
\begin{align} \label{nonuniformdyna}
\dot{x}_i=A_i x_i+B_i u_i && i=1,\dots,N,
\end{align}
with \jln{random} initial states as in \jln{Section \ref{section: fixedpointstoch}}. 
Player $i$, $i=1,\dots,N$, is associated with 
the following individual cost:
\iftoggle{jou}{
\begin{multline}	\label{eq:sysininitial}
J_{i}(u_{i},\bar{x},x_i^0) =  
\mathbb{E} \bigg ( \int_0^T \; \left \{ \frac{q}{2} \| x_{i} - \bar{x} \|^{2} + \frac{r}{2} \| u_{i} \|^{2} \right \} \,\mathrm{d}t \\ 
+ \min\limits_{j=1,\dots,l} 
\Big( \frac{M_{ij}}{2} \| x_{i}(T)-p_{j} \|^{2}   \Big) \Big | x_i^0 \bigg ),      
\end{multline}}{
\begin{equation}	\label{eq:sysininitial}
J_{i}(u_{i},\bar{x},x_i^0) =  
\mathbb{E} \bigg ( \int_0^T \; \left \{ \frac{q}{2} \| x_{i} - \bar{x} \|^{2} + \frac{r}{2} \| u_{i} \|^{2} \right \} \,\mathrm{d}t  
+ \min\limits_{j=1,\dots,l} 
\Big( \frac{M_{ij}}{2} \| x_{i}(T)-p_{j} \|^{2}   \Big) \Big | x_i^0 \bigg ).      
\end{equation}
}
%
\jln{As $N$ tends to infinity},
it is convenient to represent the limiting sequence
of $(\theta_i)_{i=1,\dots,N}=((A_i,B_i,M_{i1}, \dots ,M_{il}))_{i=1,\dots,N}$ by a random
vector $\theta$. We assume that $\theta$ is 
in a compact set $\Theta$.  
Let us denote the empirical measure of the sequence $\theta_i$ as    
$P^N_\theta(\mathcal A) = 1/N 
\sum_{i=1}^N 1_{\{\theta_i \in \mathcal A
\}}$ for 
all (Borel) measurable sets $\mathcal A$. 
We assume that $P^N_\theta$  
has a weak limit $P_\theta$, 
that is for all $\phi$ continuous,  $\lim_{N\to\infty} \int_{\Theta} 
\phi (x) 
\mathrm{d}P^N_\theta(x)=\int_{\Theta} 
\phi (x) \mathrm{d}P_\theta(x)$. 
For further discussions about this assumption, one can refer 
to \cite{huang2012social}. We assume that the initial states $x_i^0$ and $\theta$ are independent. 

In the costs \eqref{eq:sysininitial}, a small $M_{ij}$ \jln{relative to $M_{ik}$, $k \neq j$,}
reflects an a priori affinity of agent $i$ towards the destination $p_j$.
We assume that an agent $i$ knows its initial position $x_i^0$, its parameters $\theta_i$, as well as the distributions $P_0$ and $P_\theta$.
We develop the following analysis for a generic agent with an initial position $x^0$ and parameters
$\theta$. 
Assuming an infinite size population, we start by tracking $\hat{x}(t)$, a posited deterministic although initially unknown continuous path. We can then show that, \rsa{under the convention in Assumption \ref{assum-rule}}, this tracking problem is associated with  a unique optimal control law 
\begin{align}\label{eq:ctrinit}
u^{*}(t) = 
- \frac{1}{r} (B^\theta)^T \big (\Gamma_j^{\theta}(t) x + \beta_j^{\theta}(t) \big ) &&
\text{if } x^0 \in D_{j}^{\theta}(\hat{x}),
\end{align}
where $\Gamma_{j}^{\theta}$, $\beta_{j}^{\theta}$, $\delta_{j}^{\theta}$ are the unique solutions of
\begin{subequations}
\begin{align}
&\dot{\Gamma}_{j}^{\theta}-\frac{1}{r} \Gamma_{j}^{\theta} B^\theta (B^\theta)^{T}\Gamma_{j}^{\theta} +\Gamma_{j}^{\theta} A^\theta+(A^\theta)^{T}\Gamma_{j}^{\theta} +q I_{n}=0  \label{eq:abd-1init} \\
&\dot{\beta}_{j}^{\theta}=\left(\frac{1}{r} \Gamma_{j}^{\theta} B^\theta (B^\theta)^{T}-(A^\theta)^{T}\right )\beta_{j}^{\theta} +q\hat{x} \label{eq:abd-2init} \\
&\dot{\delta}_{j}^{\theta}=\frac{1}{2r} (\beta_{j}^{\theta})^{T}B^\theta (B^\theta)^{T}\beta_{j}^{\theta} -\frac{1}{2}q  {\hat{x}}^{T}\hat{x},  \label{eq:abd-3init} 
\end{align}
\end{subequations}
with the final conditions $
\Gamma_{j}^{\theta} (T)=M_{j}^\theta I_{n}, \;\; \beta_{j}^{\theta} (T)=-M^\theta_{j}p_j,
\delta_{j}^{\theta} (T)=\frac{1}{2}M^\theta_{j} p_j^{T} p_j$.
The definition of the basins of attraction becomes
\iftoggle{jou}{
\begin{multline} \label{eq:daeinit}
D_{j}^{\theta}(\hat{x}) = \Big \{ x\in \mathbb R^n \text{ such that}\\ x^T\Gamma_{jk}^{\theta}x+x^T\beta_{jk}^{\theta}(\hat{x})+\delta_{jk}^{\theta}(\hat{x}) \leq 0 ,\, \forall k=1,\dots,l \Big \}, 
\end{multline}}{
\begin{equation} \label{eq:daeinit}
D_{j}^{\theta}(\hat{x}) = \Big \{ x\in \mathbb R^n \text{ such that } x^T\Gamma_{jk}^{\theta}x+x^T\beta_{jk}^{\theta}(\hat{x})+\delta_{jk}^{\theta}(\hat{x}) \leq 0 ,\, \forall k=1,\dots,l \Big \}, 
\end{equation}
}
where
\begin{equation} \label{eq:betainit}
\begin{split}
&\Gamma_{jk}^{\theta}=\Gamma_{j}^{\theta}(0)-\Gamma_{k}^{\theta}(0)\\
&\beta_{jk}^{\theta}(\hat{x})=\beta_j^{\theta}(0)-\beta_k^{\theta}(0)\\
&\delta_{jk}^{\theta}(\hat{x})=\delta_j^{\theta}(0)-\delta_k^{\theta}(0).\\
\end{split}
\end{equation}
In this case, the solutions of the Riccati equations (\ref{eq:abd-1init}) 
depend on both the initial preference vector $M^\theta$ and the destination points. 
Hence, the basins of attraction are now regions delimited by quadric surfaces in $\mathbb{R}^n$ 
instead of hyperplanes. This fact complicates the structure of the operator that maps the 
tracked path to the mean. 
The existence proof for a fixed point relies now on an abstract Banach space version 
of Brouwer's fixed point theorem, namely Schauder's fixed point theorem \cite{FunAnaCon}. We define
\[
\Psi_j^{\theta}(\eta_1,\eta_2,\eta_3,\eta_4)=\Phi_j^\theta(\eta_1,\eta_2)^TB^\theta (B^\theta )^{T}\Phi_j^\theta(\eta_3,\eta_4),
\]
where $\Pi_{j}^{\theta}(t)=\frac{1}{r}\Gamma_j^{\theta}(t)B^\theta (B^\theta)^T-(A^\theta)^T$, 
and $\Phi_{j}^{\theta}$ is defined as in (\ref{TM}), where $\Pi$ is replaced by
$\Pi_{j}^{\theta}$. 
The state trajectory of the generic agent is then
\iftoggle{jou}{
\begin{multline*} 
x^{0\theta}(t)  =\sum_{j=1}^{l}  1_{D^{\theta}_{j}(\hat{x})}(x^0) \Big \{\Phi_{j}^{\theta}(0,t)^Tx^0 \\
+\frac{M_j^\theta}{r}\int_0^t \! \ \Psi_j^{\theta}(\sigma,t,\sigma,T)p_{j} \, \mathrm{d}\sigma  \\ 
 -\frac{q}{r}\int_0^t \! \ \int_T^\sigma \! \ \Psi_j^{\theta}(\sigma,t,\sigma,\tau)\hat{x}(\tau) \, \mathrm{d}\tau \mathrm{d}\sigma \Big \}. 
\end{multline*}}{
\begin{multline*} 
x^{0\theta}(t)  =
\sum_{j=1}^{l}  1_{D^{\theta}_{j}(\hat{x})}(x^0) \Big \{\Phi_{j}^{\theta}(0,t)^Tx^0 
+\frac{M_j^\theta}{r}\int_0^t \! \ \Psi_j^{\theta}(\sigma,t,\sigma,T)p_{j} \, \mathrm{d}\sigma \\  
 -\frac{q}{r}\int_0^t \! \ \int_T^\sigma \! \ \Psi_j^{\theta}(\sigma,t,\sigma,\tau)\hat{x}(\tau) \, \mathrm{d}\tau \mathrm{d}\sigma \Big \}. 
\end{multline*}
}

\begin{assumption}	\label{assumption: finite mean}
We assume that $\mathbb{E}\|x^0\|^2 < \infty$.
\end{assumption}

The functions defined by (\ref{eq:abd-1init}), (\ref{eq:abd-2init}) and (\ref{eq:abd-3init}) are 
continuous with respect to $\theta$ which belongs to a compact set. 
Moreover, $\theta$ and $x^0$ are assumed to be independent. 
Thus, under Assumption \ref{assumption: finite mean}, the mean of the infinite size population can be computed 
using Fubini-Tonelli's theorem \cite{Rudin87_Real} as follows:
\iftoggle{jou}{
\begin{multline}  \label{meaninit}
\bar{x}(t)=\mathbb{E}(x^{0\theta}(t))=\mathbb{E}\Big(\mathbb{E}\big(x^{0\theta}(t)|\theta\big)\Big)=\mathbb{E}\bar{x}^{\theta}(t)
\\ =\sum_{j=1}^{l} \int_\Theta \, \int_{\mathbb{R}^n} \, 1_{D^{\theta}_{j}(\hat{x})}(x^0) \Big \{\Phi_j^\theta(0,t)^Tx^0 \\
+\frac{M_j^\theta}{r}\int_0^t \! \ \Psi_j^{\theta}(\sigma,t,\sigma,T)p_{j} \, \mathrm{d}\sigma  \\ 
 -\frac{q}{r}\int_0^t \! \ \int_T^\sigma \! \ \Psi_j^{\theta}(\sigma,t,\sigma,\tau)\hat{x}(\tau) \, \mathrm{d}\tau \mathrm{d}\sigma \Big \} \, \mathrm{d}P_0\mathrm{d}P_\theta, 
\end{multline}}{
\begin{multline}  \label{meaninit}
\bar{x}(t)=\mathbb{E}(x^{0\theta}(t))=\mathbb{E}\Big(\mathbb{E}\big(x^{0\theta}(t)|\theta\big)\Big)=\mathbb{E}\bar{x}^{\theta}(t) =\sum_{j=1}^{l} \int_\Theta \, \int_{\mathbb{R}^n} \, 1_{D^{M\theta}_{j}(\hat{x})}(x^0) \Big \{\Phi_j^\theta(0,t)^Tx^0 \\
+\frac{M_j^\theta}{r}\int_0^t \! \ \Psi_j^{\theta}(\sigma,t,\sigma,T)p_{j} \, \mathrm{d}\sigma   
 -\frac{q}{r}\int_0^t \! \ \int_T^\sigma \! \ \Psi_j^{\theta}(\sigma,t,\sigma,\tau)\hat{x}(\tau) \, \mathrm{d}\tau \mathrm{d}\sigma \Big \} \, \mathrm{d}P_0\mathrm{d}P_\theta, 
\end{multline}
}
where $\bar{x}^{\theta}(t)=\mathbb{E}\big(x^{0\theta}(t)|\theta\big)$.
Equation (\ref{meaninit}) defines an operator $G_p$ from the Banach space $(C([0,T],\mathbb{R}^n),\|.\|_\infty)$ into itself which maps the infinite population tracked path $\hat{x}$ to the corresponding mean $\bar{x}$, itself considered as another potential tracked path. 

In the next theorem, we show that $G_p$ has a fixed point.
We define 
\rsa{\begin{equation} \label{bounds}
\begin{split}
k_1&=\mathbb{E}\|x^0\|\times\left(\sum_{j=1}^l \max\limits_{(\theta,t)\in \Theta \times 
[0,T]}\|\Phi_j^\theta(0,t)\|\right)\\
k_2&=\sum_{j=1}^l\max\limits_{(\theta,t)\in \Theta \times [0,T]}
\bigg \|\frac{M_j^\theta}{r}\int_0^t \! \ \Psi_j^{\theta}(\sigma,t,\sigma,T)p_{j} \, \mathrm{d}\sigma \bigg \| \\ 
k_3&=\frac{q}{r}\sum_{j=1}^l\max\limits_{(\theta,t,\sigma,\tau)\in \Theta \times 
[0,T]^3}\|\Psi_j^{\theta}(\sigma,t,\sigma,\tau)\|.
\end{split}
\end{equation}}
Since $\Theta$ and $[0,T]$ are compact and $\Phi_j^\theta$ is continuous
with respect to time and parameter $\theta$, then $k_1$, $k_2$ and $k_3$ are 
well defined.

\rsa{\begin{assumption}	\label{assumption: bounds}
We assume that $\sqrt{\max(k_1+k_2,k_3)}T<\pi/2$. 
\end{assumption}}
\rsa{Noting that the left hand side of the inequality tends to zero as $T$ goes to zero, Assumption \ref{assumption: bounds} can be satisfied for short time horizon $T$ for example.}
\begin{assumption}	\label{assumption: measure support}
We assume that $P_0$ is such that the $P_0$-measure of quadric surfaces is zero. 
\end{assumption}
\begin{thm} \label{lemma:existanceinit}
Under Assumptions \ref{assumption: finite mean}, \rsa{\ref{assumption: bounds}} and \ref{assumption: measure support}, 
$G_p$ has a fixed point.
\end{thm}
\begin{IEEEproof}
See Appendix \ref{Proof2}.
\end{IEEEproof}


\section{Nash Equilibrium} \label{section:Nash}

In the three cases above, deterministic, stochastic and stochastic with initial preferences, 
we defined three maps $G$, $G_s$ and $G_p$ respectively. Depending on the structure
of the game, each player can anticipate the macroscopic behavior of the limiting 
population by computing a fixed point $\hat{x}$ of $G$, $G_s$ or $G_p$, and  
compute its best response $u_i^*(x_i,\hat{x})$ to $\hat{x}$ as defined in
\eqref{eq:ctr}, \eqref{eq:ctrinit}. When considering the finite population, the next theorem 
establishes the importance of such  decentralized strategies in that they lead to an 
$\epsilon$-Nash equilibrium  with respect to the costs \eqref{eq:sysinde}, \eqref{eq:sysin}
and \eqref{eq:sysininitial}. This equilibrium makes the group's behavior robust in the 
face of potential selfish behaviors as unilateral deviations 
from the associated control policies are guaranteed to yield negligible cost reductions 
as $N$ increases sufficiently.
\begin{thm} \label{theorem:nash} Under Assumption \ref{assumption: finite mean}, 
the decentralized strategies $u_i^*$, $i=1,\dots,N$, defined in (\ref{eq:ctr}) 
and (\ref{eq:ctrinit}) for a fixed point path
$\hat{x}$, constitute an $\epsilon_N$-Nash 
equilibrium with respect to the costs $J_{i}(u_i,u_{-i})$, where $\epsilon_N$ goes 
to zero as $N$ increases to infinity.
\end{thm} 

\begin{IEEEproof}
See Appendix \ref{Proof3}.
\end{IEEEproof}

\section{Simulation Results} \label{section: simulation}

To illustrate the collective decision-making mechanism, we consider a group 
of agents moving in $\mathbb{R}^2$ according to the dynamics
\begin{align*}
A=\begin{bmatrix}
0 && 1\\
0.02 && -0.3
\end{bmatrix} &&
B=\begin{bmatrix}
0\\
0.3
\end{bmatrix}
\end{align*}
towards the potential destination points $p_{1}=(-39.3,-10)$, $p_{2}=(-27,9.5)$  or $p_{3}=(0,40)$.
We draw $N=600$ initial conditions from the Gaussian distribution 
$\jln{P_0} := \mathcal{N}\left(\begin{bmatrix}-10 & 0\end{bmatrix}^T,5 \, I_2\right)$. 
We simulate two cases. In the first one, each agent knows the exact  initial states 
of the other agents and anticipates the mean of the population accordingly. Following 
the counting and verification operations described at the end of Section 
\ref{section: fixed}, we find that $F$ has multiple fixed points,
for example, $\lambda=(564,11,25)$.
By implementing the control laws corresponding to this particular $\lambda$, 
$564$ agents go towards $p_1$, $11$ towards $p_2$ and the rest towards $p_3$ 
(see Fig.\ref{Fig.1}). Moreover, the actual average replicates the anticipated mean 
as shown in this figure. In the second case, the agents know only the initial distribution 
\jln{$P_0$} of the agents. 
Then, Broyden's method converges to $\lambda = (0.9162,0.0258,0.058)$ 
satisfying (\ref{stineq-multi}). Accordingly, $91.62\%$ of the agents go towards 
$p_1$, $2.58\%$ towards $p_2$ and the rest towards $p_3$ 
(see Fig.\ref{Fig.2}). 
The actual average and the anticipated mean are approximately the same.

To illustrate the social effect
on the individual choices (see Fig. \ref{Fig.3}), we consider the same initial conditions. 
Without social effect ($q=0$), $(0,0.25,0.75)$ satisfies (\ref{stineq-multi}). In this case, 
the majority goes towards $p_3$. As the social effect increases to $q=4$, some of 
the agents that went towards $p_1$ or $p_2$ in the absence of a social effect 
change their decisions and follow the majority towards $p_3$ 
(see blue balls in Fig. \ref{Fig.3}). In this case, $(0,0.16,0.84)$ satisfies 
(\ref{stineq-multi}). If the social impact increases more to $q=6$, then a consensus 
to follow the majority occurs. 

To illustrate the impact of the individual efforts on the behavior of the population (see Fig. \ref{Fig.4}), 
we start with the case where control effort is inexpensive ($r=3$) relative  to the social effect ($q=14$). In this case, each agent prefers following 
the majority. A consensus to go towards $p_3$ occurs. 
As the effort coefficient increases ($r=30$), some of the agents prefer going 
to a less expensive destination ($p_1$) than following the majority. By increasing more the penalty on the effort ($r=60$), a third party appears. This subgroup goes towards $p_2$. 
Moreover, when $r$ decreases, the agents reach 
smaller neighborhoods of the destination points. 

To illustrate the Gaussian Binary Choice Case, we consider a population of $N=500$ agents 
initially drawn from the normal distribution $\mathcal{N}(\mu_0,15I_2)$ and moving 
in $\mathbb{R}^2$ according to the dynamics $A=B=I_2$ towards the destination points 
$p_{1}=(-20,0)$ or $p_{2}=(20,0)$. For this covariance matrix $\Sigma_0=15 I_2$, $S(\Sigma_0)$ 
is the region delimited by the vertical lines $x=-15$ and $x=15$. If $\mu_0=(18\,\,5)$, i.e. 
outside $S(\Sigma_0)$, only one $\epsilon-$Nash equilibrium corresponding 
to $\lambda=(0,1)$ exists.
If $\mu_0=(0.5\,\,5)$, i.e. inside $S(\Sigma_0)$, three $\epsilon$-Nash equilibria exist. The first corresponds to $\lambda_1=(0.89,0.11)$ (Fig. \ref{Fig.9}), the second 
to $\lambda_2=(0.4,0.6)$ (Fig. \ref{Fig.10}), and the third to $\lambda_3=(0,1)$ (Fig. \ref{Fig.11}).

\iftoggle{jou}{

\begin{figure}[h!]
    \centering
    \includegraphics[width=0.49\textwidth]{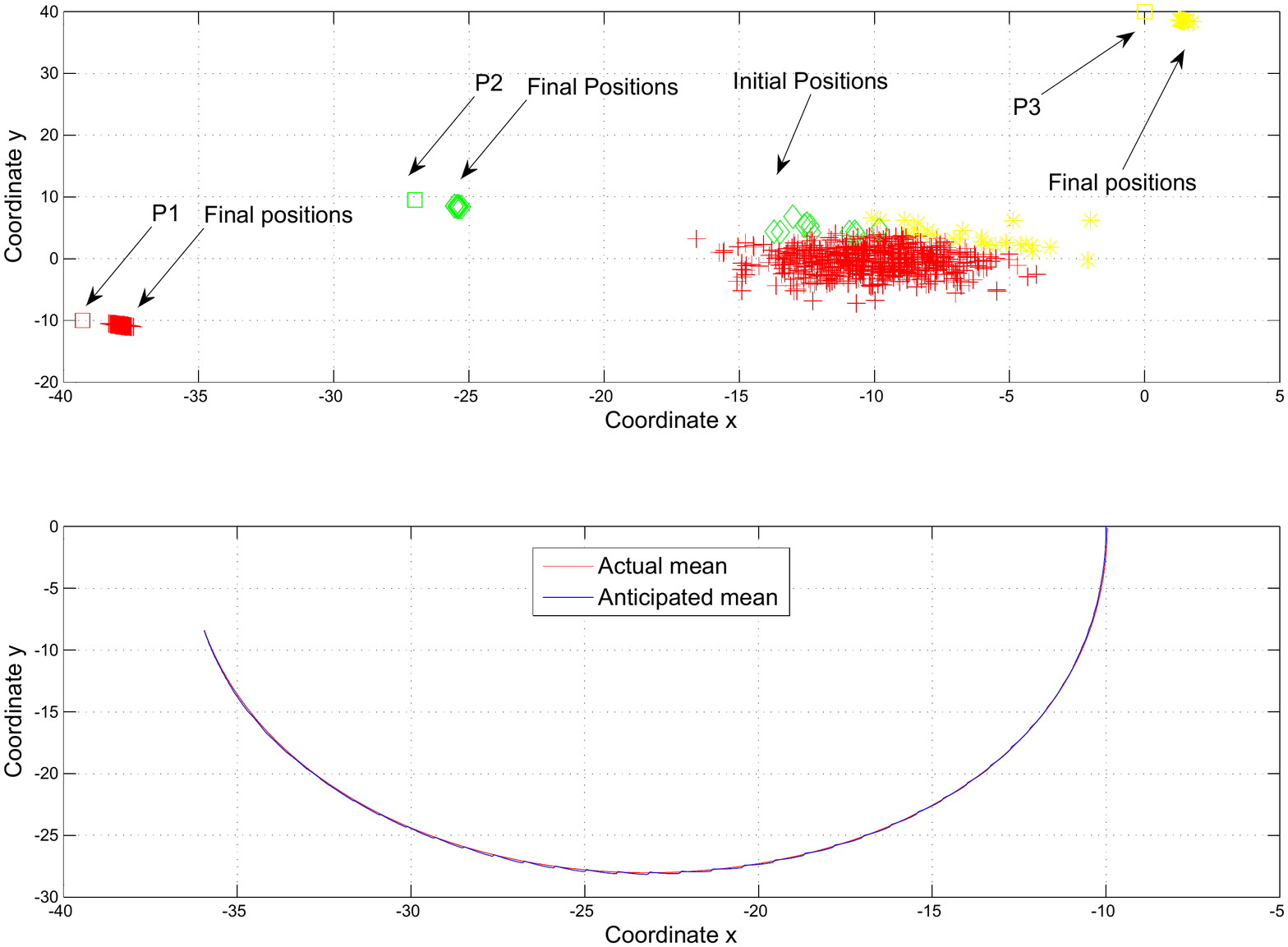}
    \caption{Collective choice - deterministic initial conditions - $\lambda = (564,11,25)$}
    \label{Fig.1}
\end{figure}

\begin{figure}[h!]
    \centering
    \includegraphics[width=0.49\textwidth]{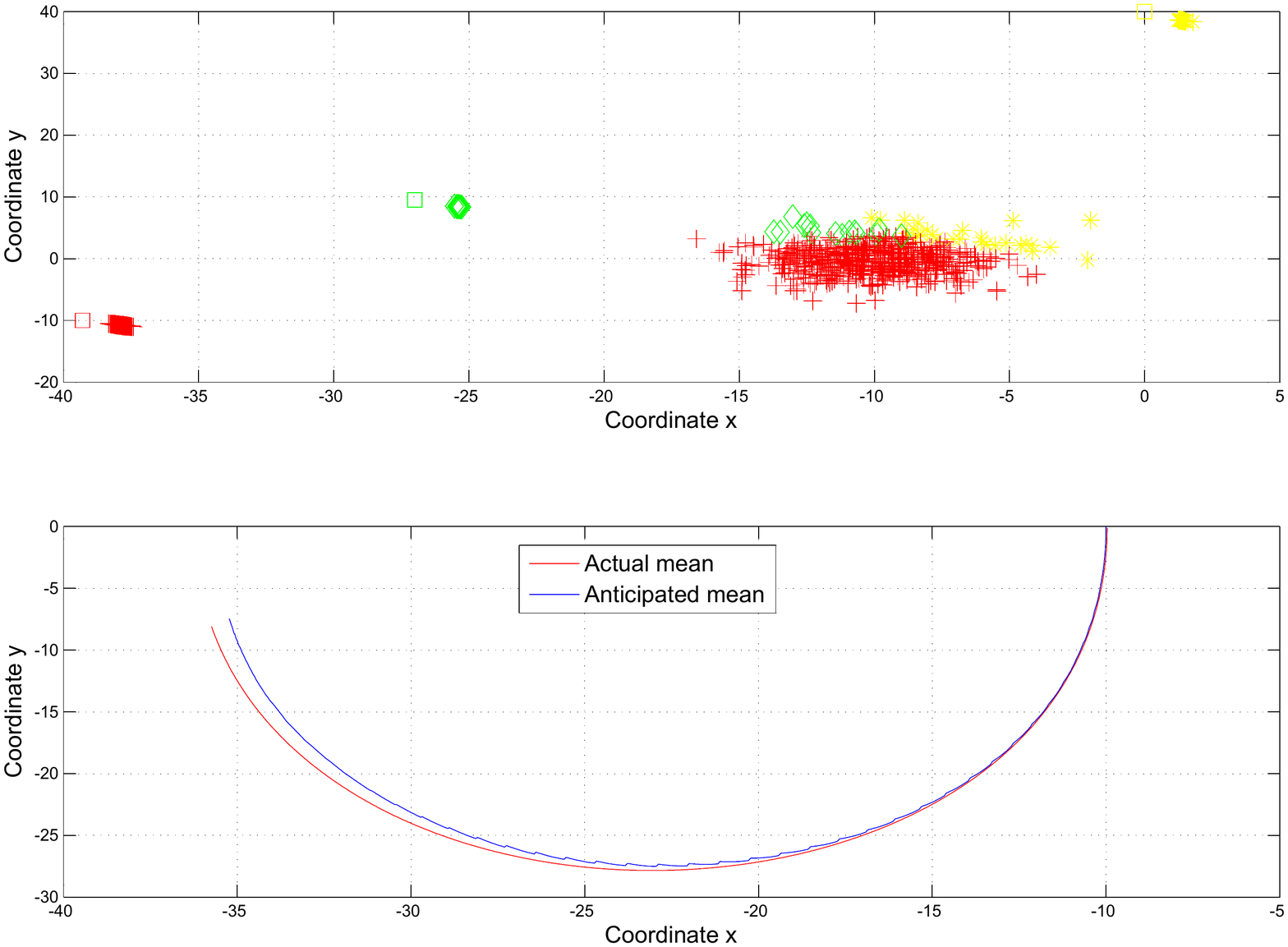}
    \caption{Collective choice - random initial conditions - $\lambda = (0.9162,0.0258,0.058)$}
    \label{Fig.2}
\end{figure}

\begin{figure}[h!]
    \centering
    \includegraphics[width=0.5\textwidth]{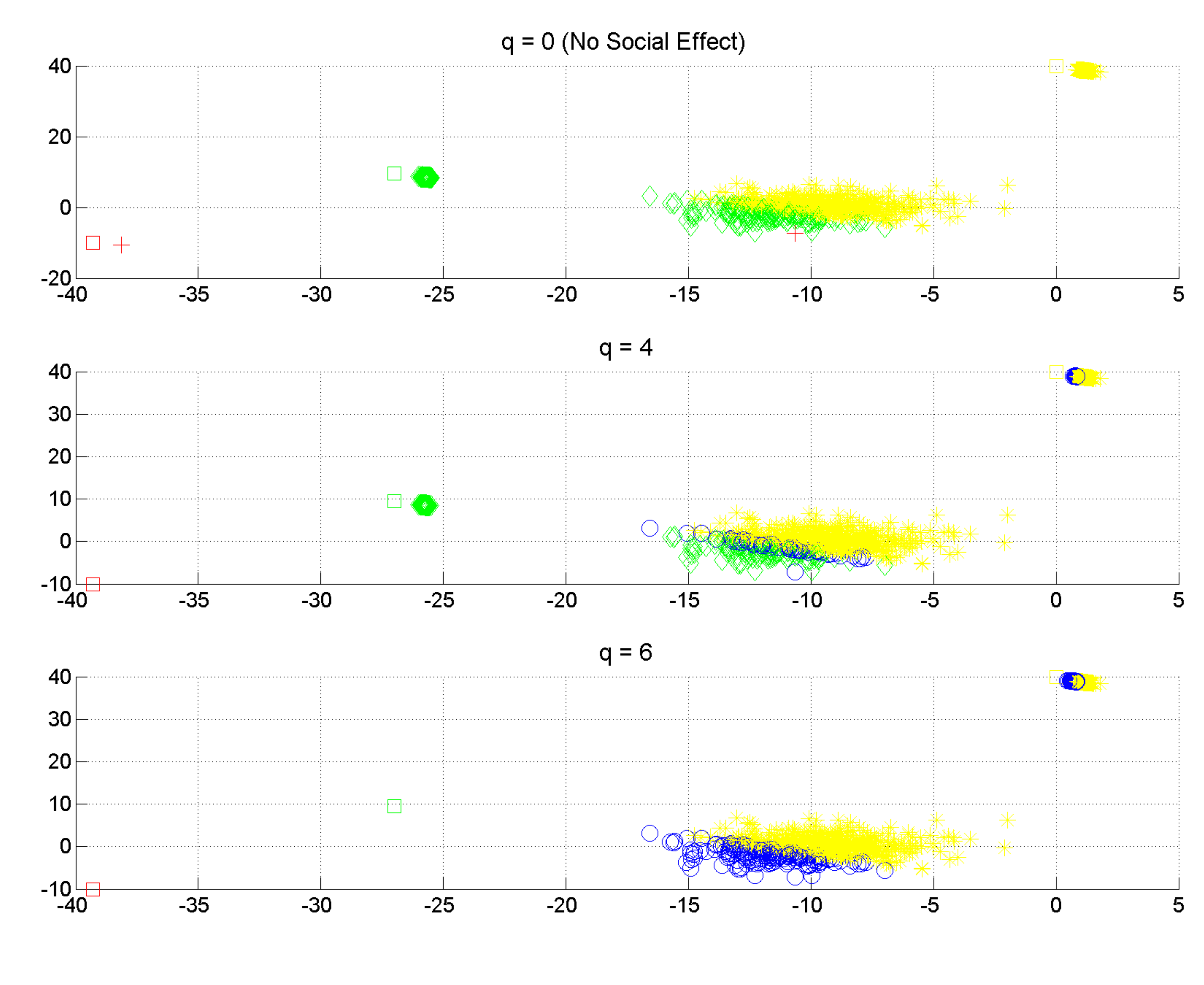}
    \caption{Influence of the social effect $q$}
    \label{Fig.3}
\end{figure}

\begin{figure}[h!]
    \centering
    \includegraphics[width=0.5\textwidth]{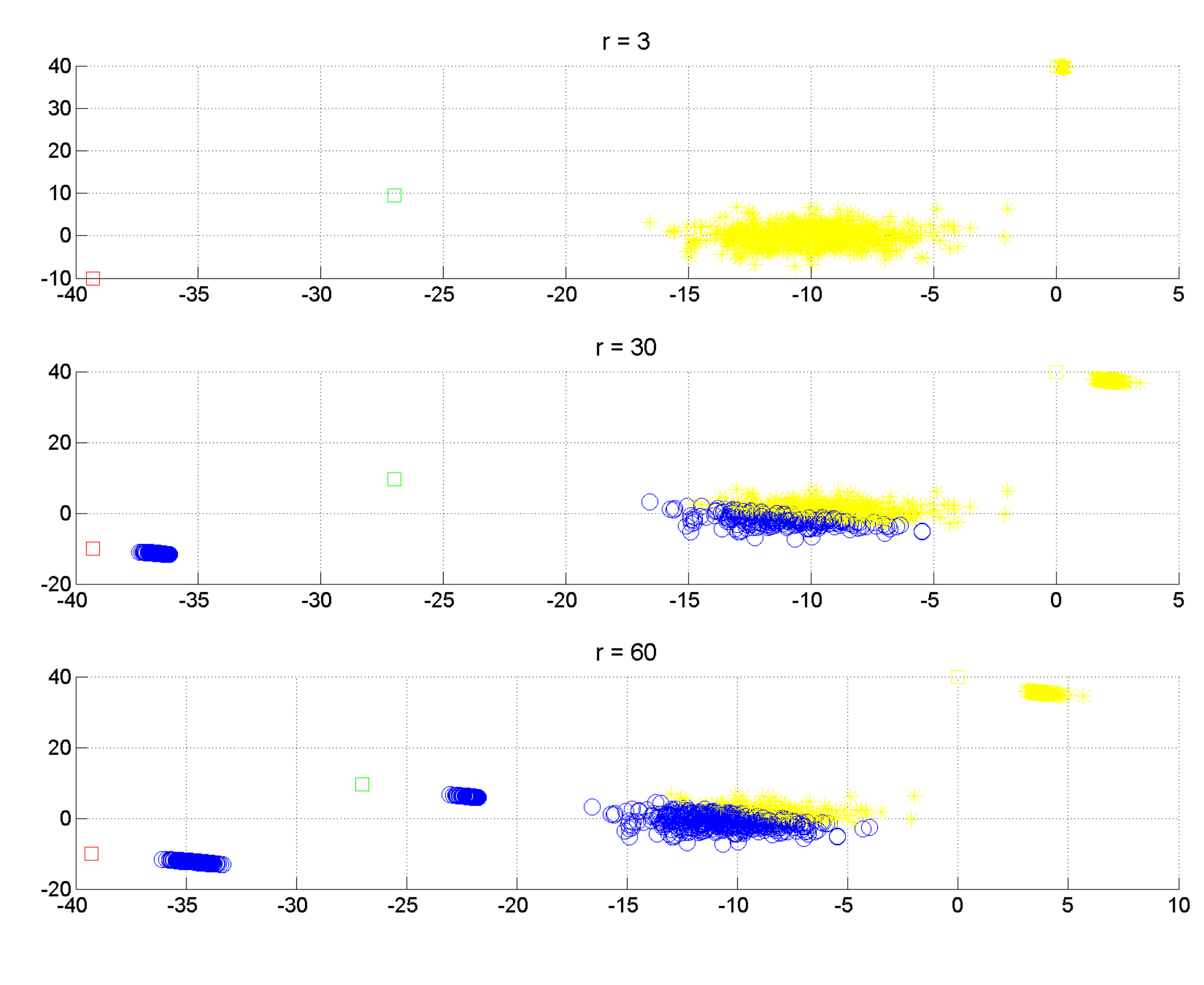}
    \caption{Influence of $r$}
    \label{Fig.4}
\end{figure}
\begin{figure}[h!]
    \centering
\includegraphics[width=0.5\textwidth]{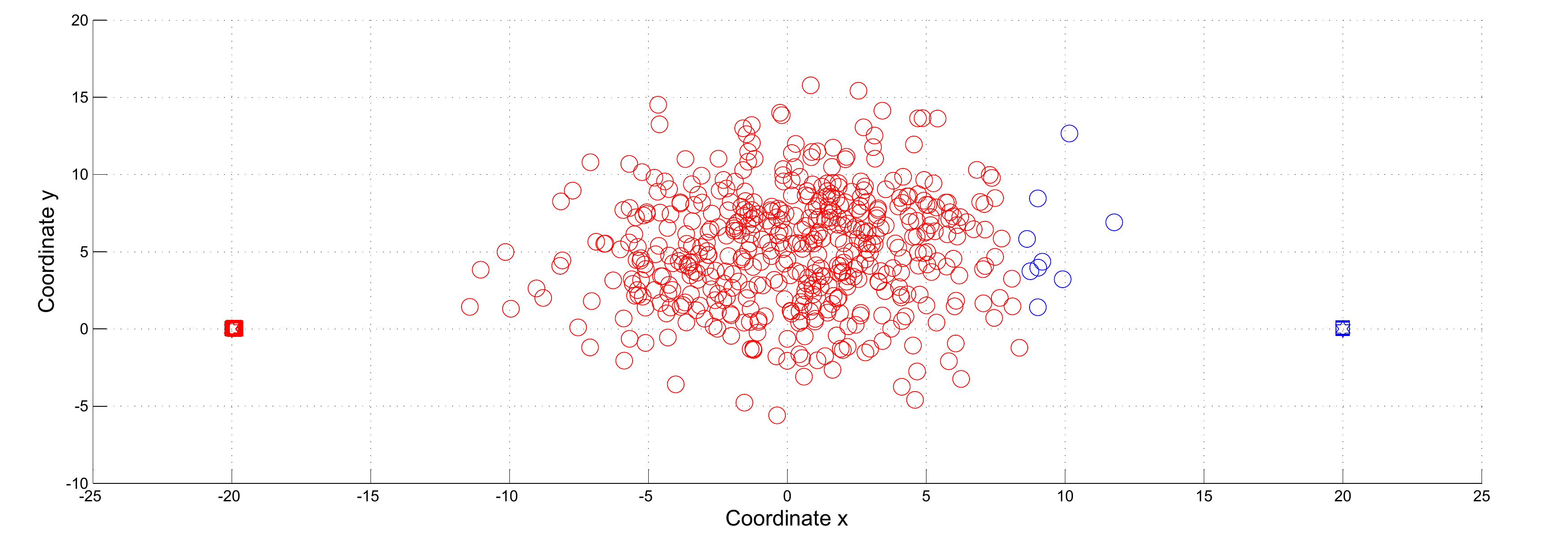}
    \caption{Gaussian binary choice case - Multiple equilibria - $\lambda = (0.89,0.11)$}
    \label{Fig.9}
\end{figure}
\begin{figure}[h!]
    \centering
\includegraphics[width=0.5\textwidth]{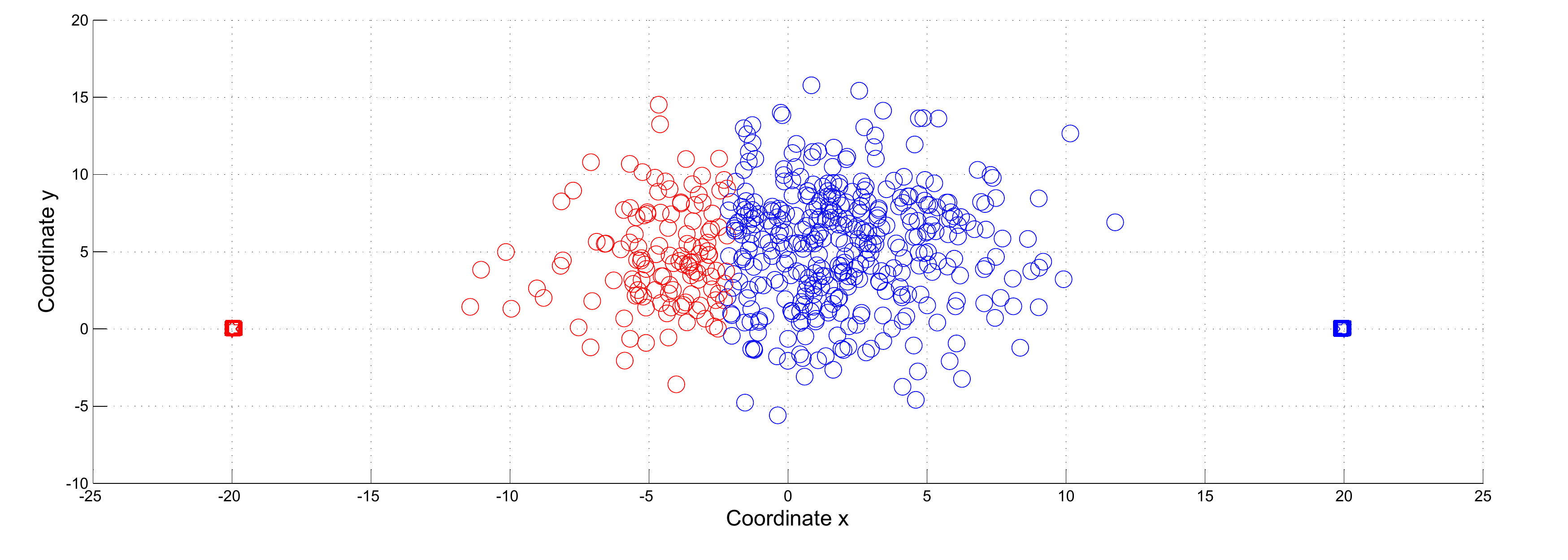}
    \caption{Gaussian binary choice case - Multiple equilibria - $\lambda = (0.4,0.6)$}
    \label{Fig.10}
\end{figure}
\begin{figure}[h!]
    \centering
\includegraphics[width=0.5\textwidth]{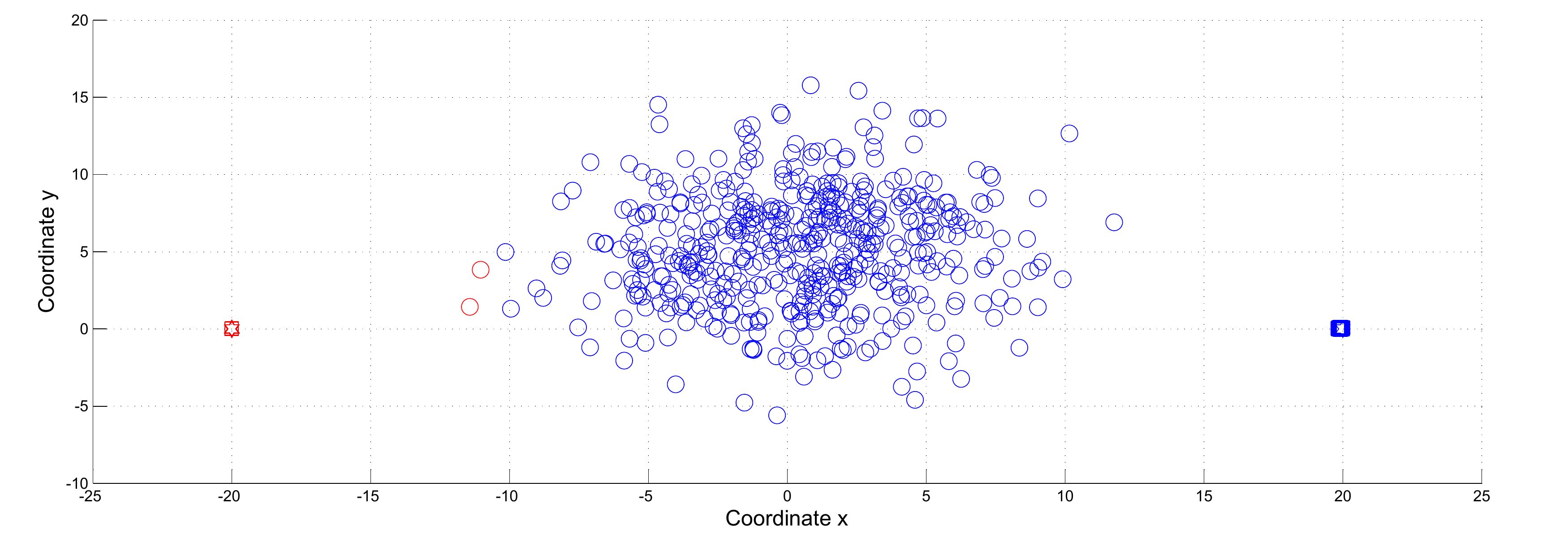}
    \caption{Gaussian binary choice case - Multiple equilibria - $\lambda = (0,1)$}
    \label{Fig.11}
\end{figure}}{
\begin{figure}[h!]
    \centering
 \includegraphics[width=0.7\textwidth]{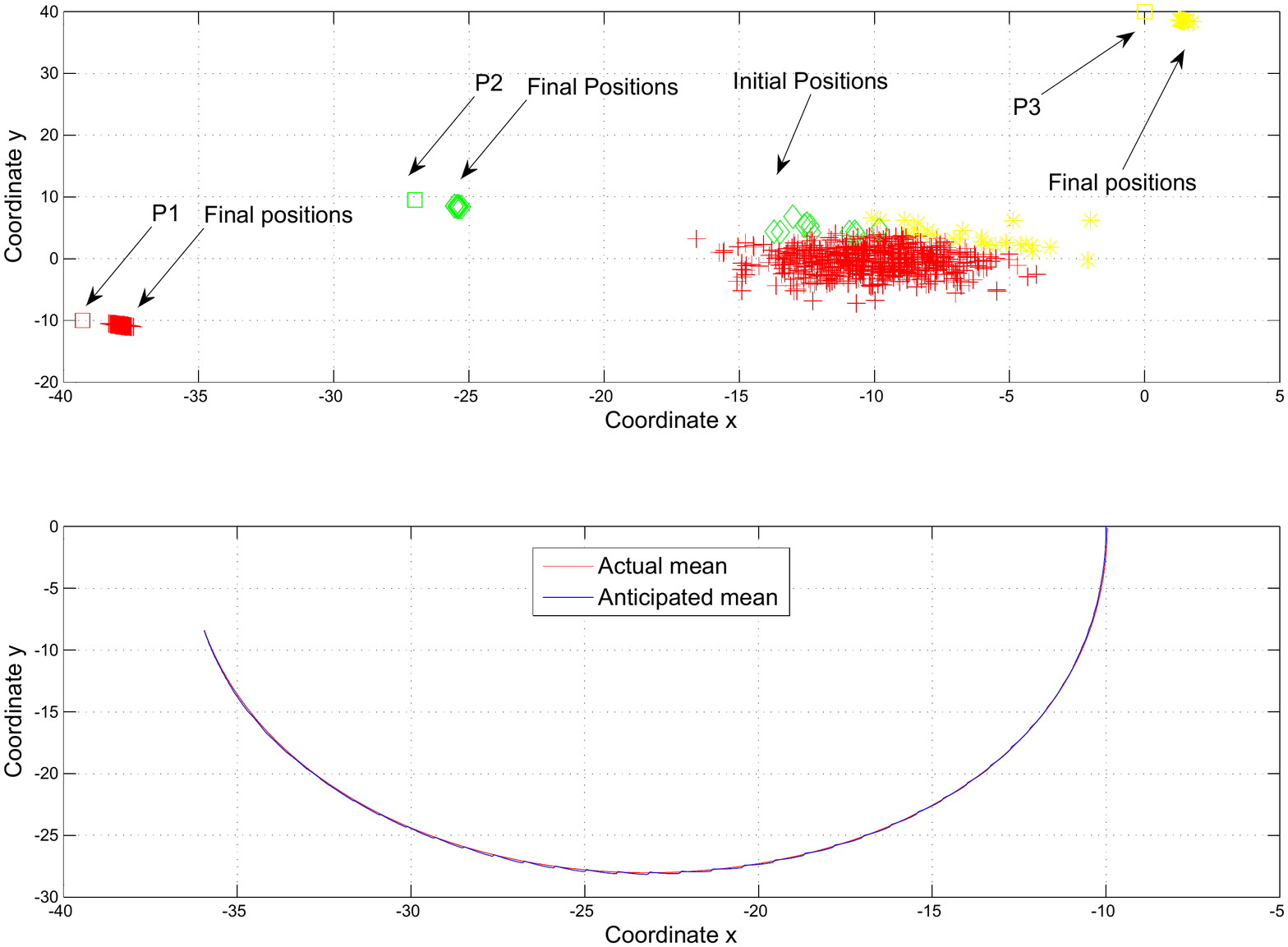}
    \caption{Collective choice - deterministic initial conditions - $\lambda = (564,11,25)$}
    \label{Fig.1}
\end{figure}

\begin{figure}[h!]
    \centering
    \includegraphics[width=0.7\textwidth]{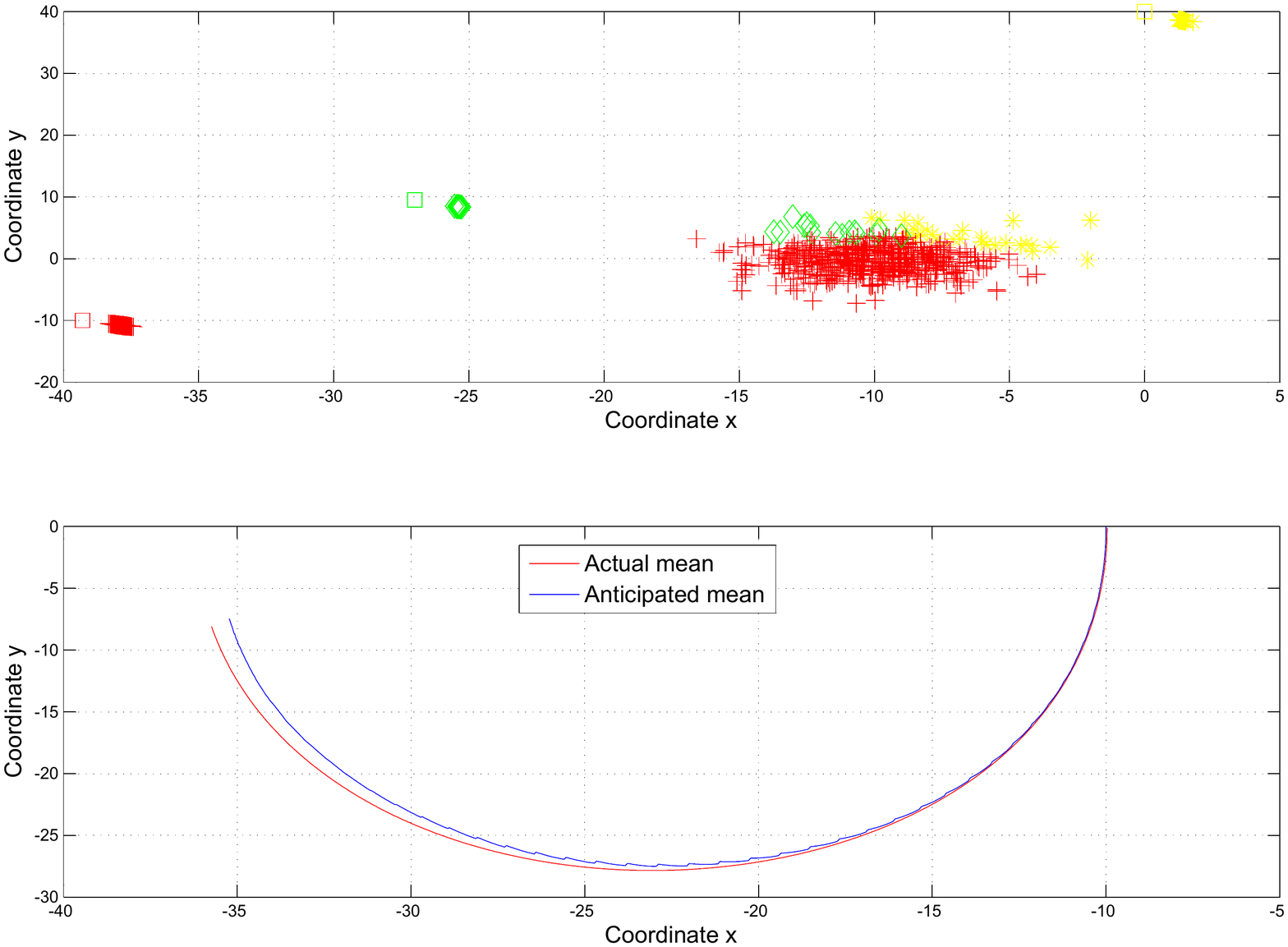}
    \caption{Collective choice - random initial conditions - $\lambda = (0.9162,0.0258,0.058)$}
    \label{Fig.2}
\end{figure}

\begin{figure}[h!]
    \centering
    \includegraphics[width=0.71\textwidth]{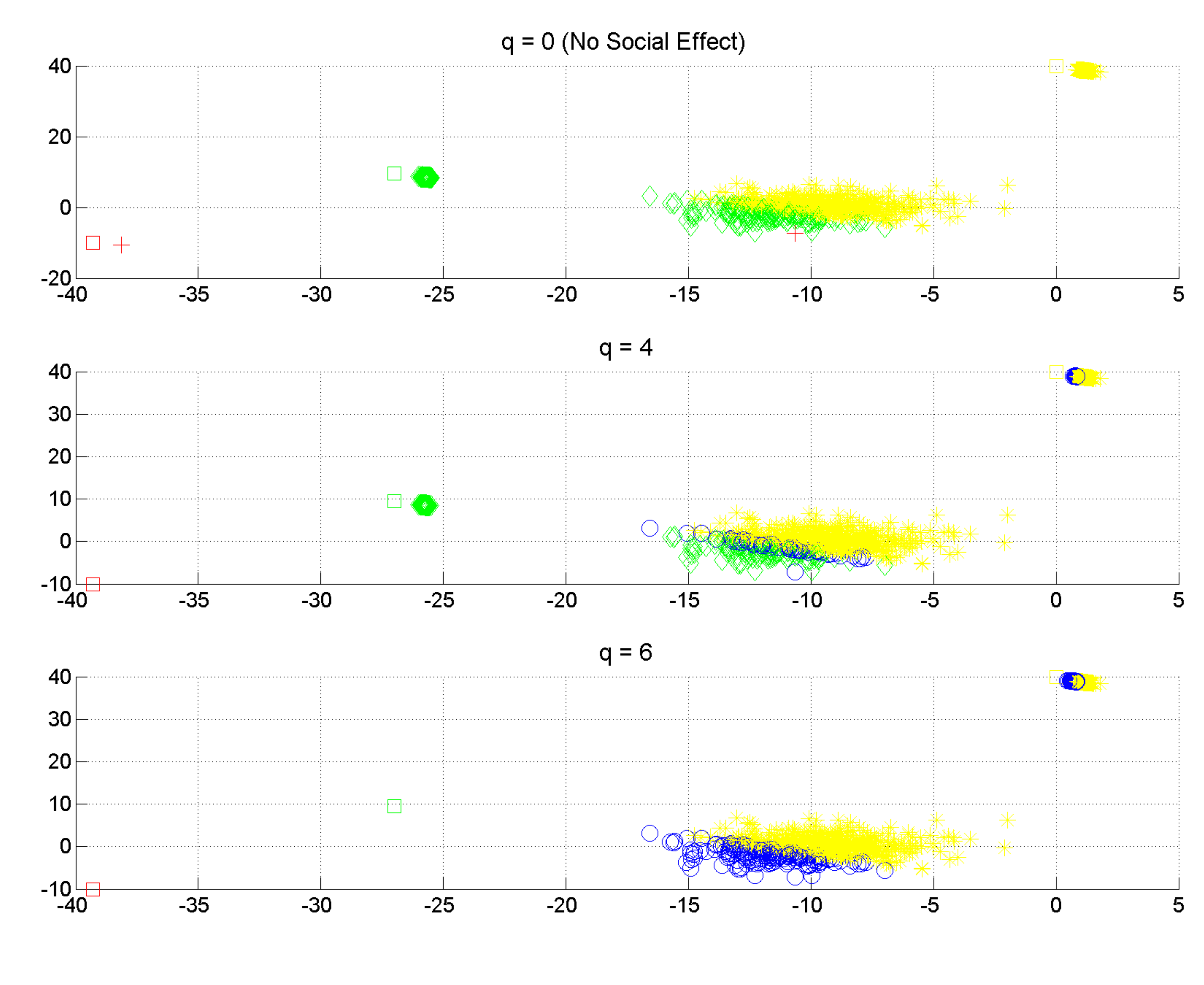}
    \caption{Influence of the social 
    effect $q$}
    \label{Fig.3}
\end{figure}

\begin{figure}[h!]
    \centering
    \includegraphics[width=0.71\textwidth]{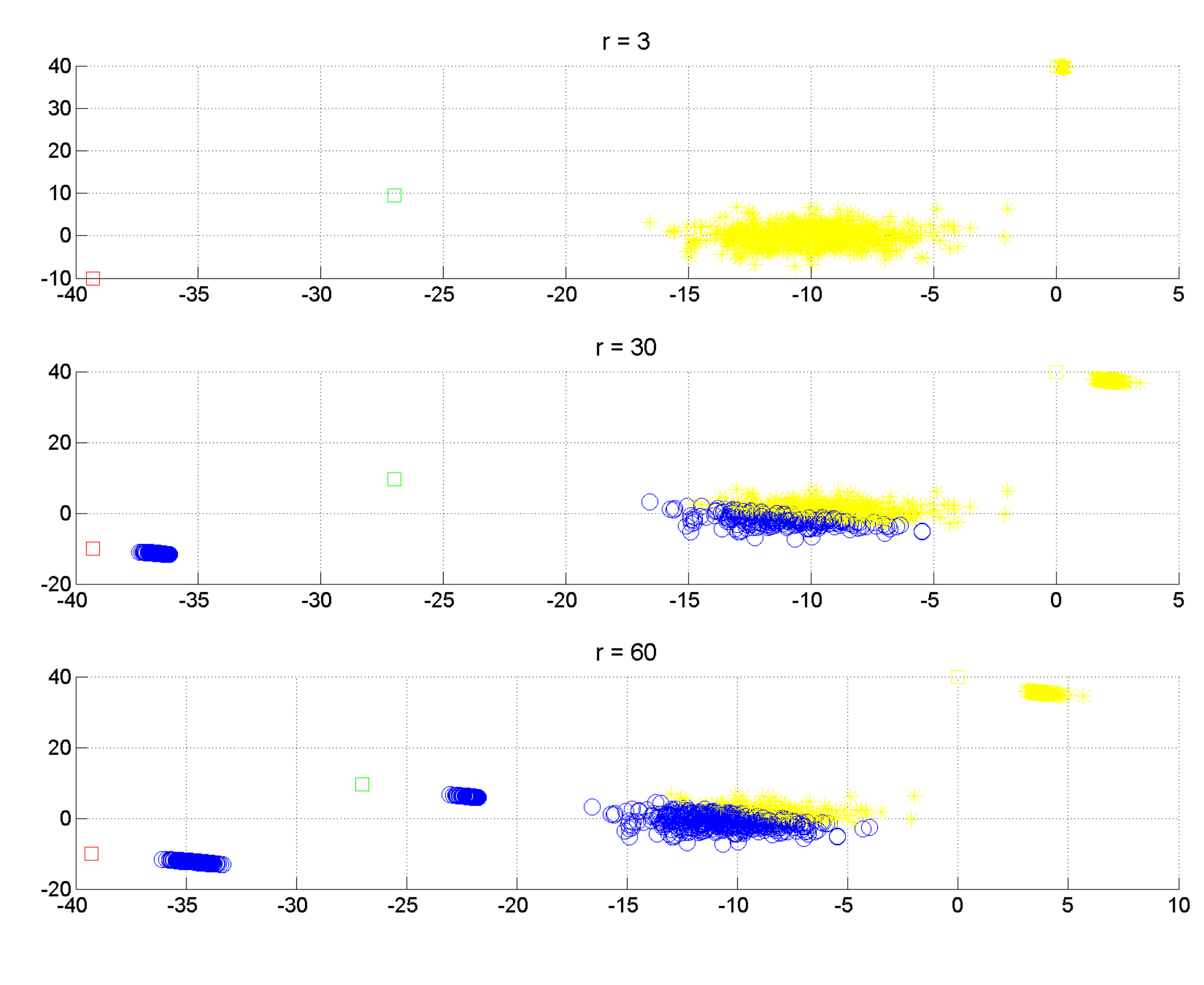}
    \caption{Influence of $r$}
    \label{Fig.4}
\end{figure}
\begin{figure}[h!]
    \centering
\includegraphics[width=0.71\textwidth]{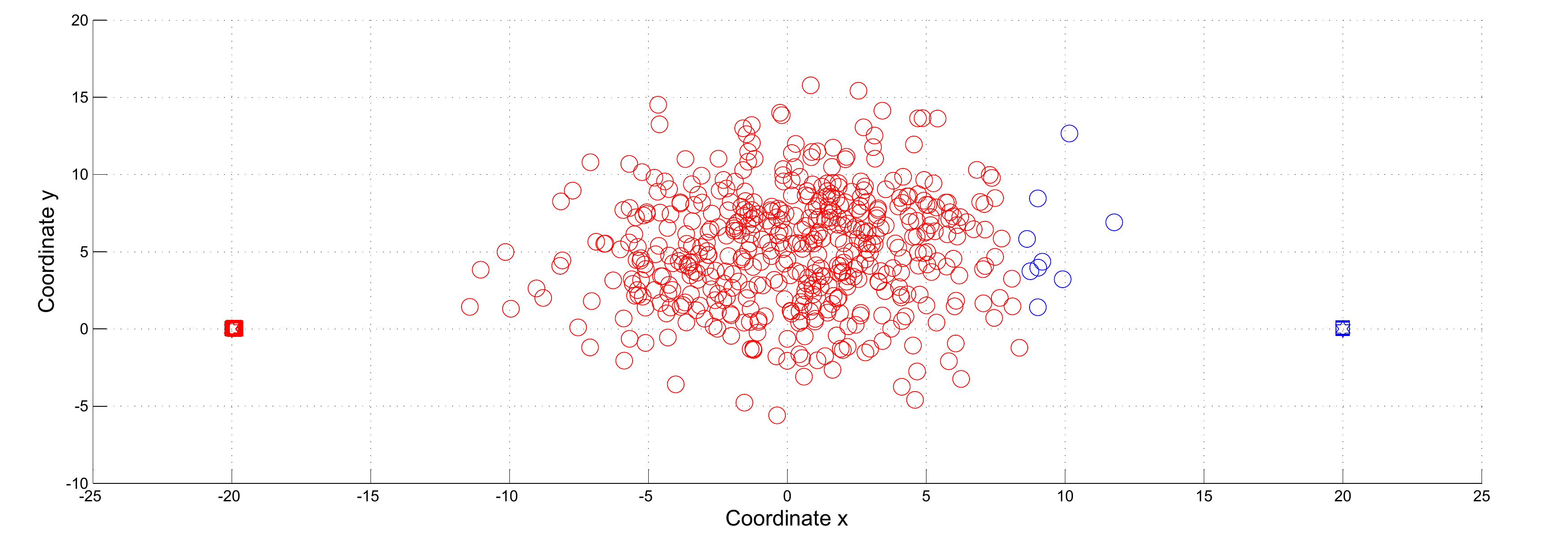}
    \caption{Gaussian binary choice case - Multiple equilibria - $\lambda = (0.89,0.11)$}
    \label{Fig.9}
\end{figure}
\begin{figure}[h!]
    \centering
\includegraphics[width=0.71\textwidth]{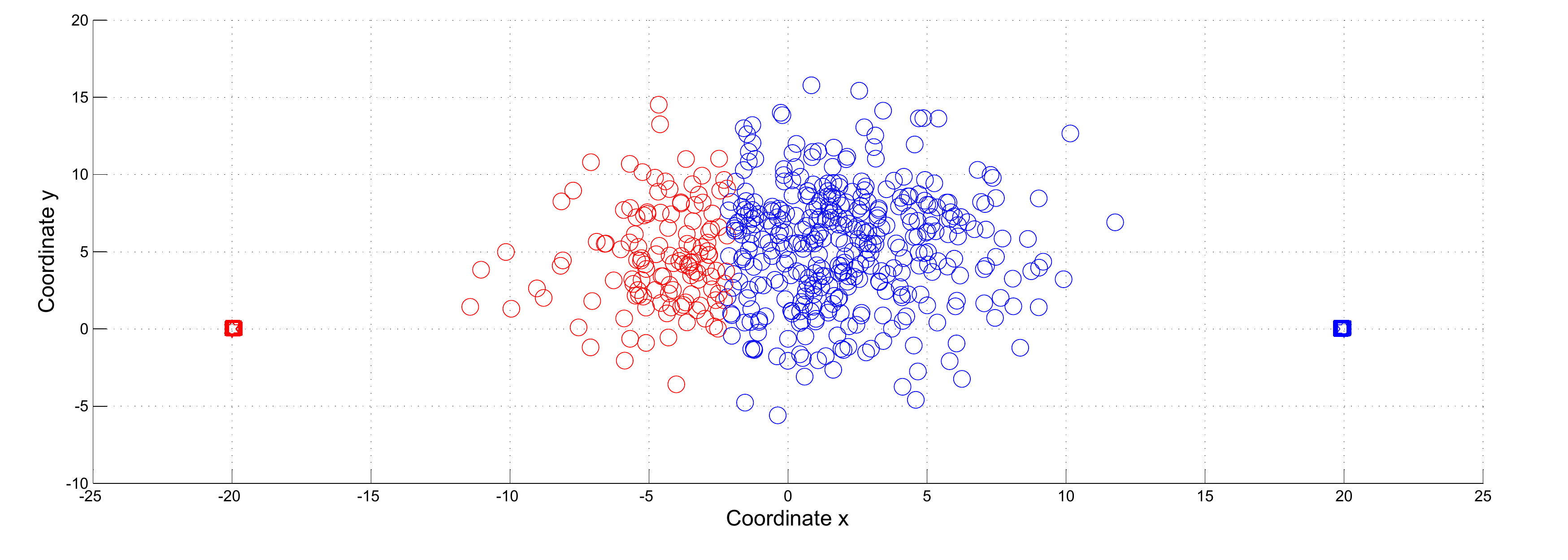}
    \caption{Gaussian binary choice case - Multiple equilibria - $\lambda = (0.4,0.6)$}
    \label{Fig.10}
\end{figure}
\begin{figure}[h!]
    \centering
\includegraphics[width=0.71\textwidth]{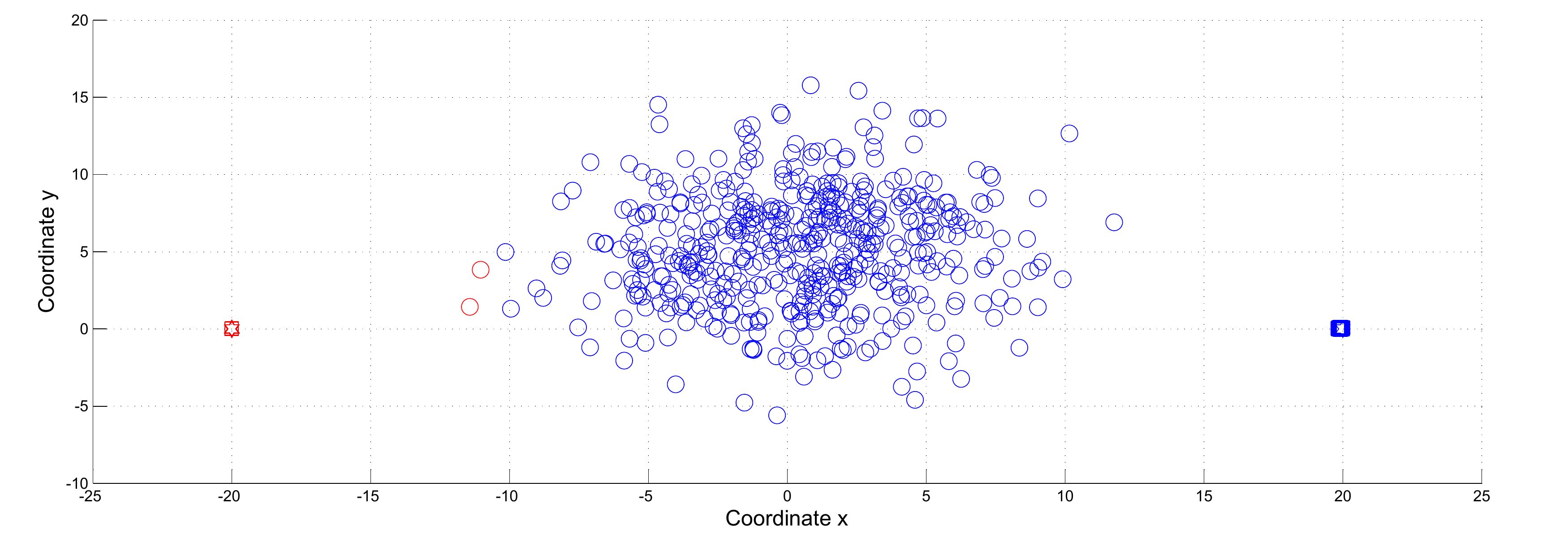}
    \caption{Gaussian binary choice case - Multiple equilibria - $\lambda = (0,1)$}
    \label{Fig.11}
\end{figure}
}

\section{Conclusion} \label{section: conclusion}
We consider in this paper a dynamic collective choice model where a large number of agents 
are choosing between multiple destination points while taking into account the social effect 
as represented by the mean of the population. \jln{The analysis is carried using the MFG methodology.}
We show that under this social effect, the population may split between the destination points in different ways.  
For a uniform population, we show that there exists a one to one map between the fixed point behaviors 
(anticipated behaviors) and the fixed points of an operator defined on $\mathbb{R}^l$. The latter 
describe the way the agents split between the $l$ destination points. Finally, we prove that 
the decentralized strategies developed while tracking the anticipated behaviors are approximate 
Nash equilibria. 
\rsa{For future work, it is of interest to analyze a 
model where stochasticity is extended to the players' dynamics as well.
In that case, the optimal choices (feedback strategies) are adapted to the underlying filtration 
and change along the path. This is in contrast to the current formulation where the agents can choose 
without loss of optimality their destination before they start moving.
Moreover, we would like to extend the current formulation to certain nonlinear models,
where the basins of attraction are delimited by more complex manifolds, and the fixed-point
computations would require numerical methods for backward-forward systems of partial differential equations 
\cite{achdou2010mean}.}


\appendices

\section{} \label{Proof1}

\subsection{Proof of Lemma \ref{contr}} 

\rsa{In this proof, the subscript $M$ 
indicates the dependence on the final cost's
coefficient $M$.}
For any $M>0$, the agents are optimally tracking a path $\hat{x}_M$.
The agent's $i$ optimal state is denoted by $x_{iM}^*(t)$. 
We have
\[
\frac{M}{2} \min\limits_{j=1,\dots,l} \Big(  \| x^*_{iM}(T)-p_{j} \|^{2}   \Big ) \leq J_{iM}(u^*_{iM},\hat{x}_M,x_i^0),
\]
where $J_{iM}(u^*_{iM},\hat{x}_M,x_i^0)$ is the cost define by (\ref{eq:sysinde})
with the final cost's coefficient equal $M$. It suffices to find an upper bound for $J_{iM}(u^*_{iM},\hat{x}_M,x_i^0)$ which is uniformly bounded with $M$. Since $(A,B)$ is controllable, then there exists for each agent $i$ a continuous control law $u_{x_i^0,p_1}(t)$ on $[0,T]$ which transfers this agent from the state $x_i^0$ to $p_1$ in a finite time $T$. 
By optimality, we have
\[
J_{iM}(u^*_{iM},\hat{x}_M,x_i^0)
\leq J_{iM}\Big(u_{x_i^0,p_1},\hat{x}_M,x_i^0 \Big). \]
But,
\iftoggle{jou}{
\begin{multline*}
J_{iM}\Big(u_{x_i^0,p_1},\hat{x}_M,x_i^0 \Big)=\\
\int_0^T \! \Big \{ \frac{q}{2} \| x_{i}({u_{x_i^0,p_1}}) -\hat{x}_M \|^{2} + 
\frac{r}{2} \| u_{x_i^0,p_1} \|^{2} \Big \} \,\mathrm{d}t,
\end{multline*}}{
\begin{equation*}
J_{iM}\Big(u_{x_i^0,p_1},\hat{x}_M,x_i^0 \Big)=
\int_0^T \! \Big \{ \frac{q}{2} \| x_{i}({u_{x_i^0,p_1}}) - \hat{x}_M \|^{2} + 
\frac{r}{2} \| u_{x_i^0,p_1} \|^{2} \Big \} \,\mathrm{d}t,
\end{equation*}
}
which is uniformly bounded with $M$, since $\hat{x}_M$ is uniformly 
bounded with $M$. Thus, for all $\epsilon>0$, there exists an $M_0>0$ such that for all $M>M_0$, 
\[
\min\limits_{j=1,\dots,l} \Big(  \| x^*_{iM}(T)-p_{j} \|^{2}   \Big )<\epsilon.
\]

\subsection{Proof of Lemma \ref{lemma:existance1}} 
Lets 
consider $y$ a fixed point of 
$T_\lambda$. We define
\begin{align*}
n(t)=\Gamma(t) y(t) + q \int_T^t \Phi 
(t,\sigma) y(\sigma) -M\Phi 
(t,T)p_\lambda.
\end{align*}
One can easily check that $(y,n)$ satisfies
\begin{align} \label{f-b}
\dot{y}&=Ay-\frac{1}{r}BB^T n &&
y(0)=\bar{x}_0 \\
\dot{n}&=-A^Tn && n(T)=M(y(T)-p_\lambda). \nonumber
\end{align}
$y$ and $n$ are respectively the optimal
state and co-state of the
following LQR problem: 
\begin{align} \label{opt_state_y}
&\min_u \int_0^T \frac{r}{2}\|u\|^2 
\mathrm{dt} + \frac{M}{2}\|x(T)-p_\lambda\|^2
\\
&\textrm{Subject to } \dot{x}=Ax+Bu && 
x(0)=\bar{x}_0. \nonumber
\end{align} 
Therefore, $n$ has the representation
$n(t)=P(t)y(t)+g(t)$, where $P$ is the 
unique solution of the 
Riccati equation (\ref{aux_riccati}) and
$g$ satisfies
\begin{align*}
\dot{g}=-(A-\frac{1}{r}BB^TP(t))^Tg &&
g(T)=-Mp_\lambda.
\end{align*}
By solving $g$ and implementing its
expression in $n=Py+g$, and by 
implementing the new expression of $n$ in
the dynamics of $y$, one can show that
$y(t)=R_1(t)\bar{x}_0+R_2(t)p_\lambda$.
Conversely, let $(n,y)$ the unique 
solution of (\ref{f-b}). We define
$m(t)=\Gamma(t)y(t)+q
\int_T^t\Phi(t,\sigma)y(\sigma)\mathrm{d\sigma}-M\Phi(t,T)p_{\lambda}$. One can 
easily check that $\dot{(m-n)}=(\frac{1}{r}\Gamma BB^T - A^T)(m-n)$, with $m(T)=n(T)$. Therefore, $m=n$. Hence, $y$
is a fixed point of (\ref{eq:muapp1}). 
We now prove the uniform boundedness of the fixed point paths $y_\lambda$
with respect to $M$. The paths $y_\lambda$ are the optimal states of the
control problem (\ref{opt_state_y}). Since $(A,B)$ is controllable, one 
can show that the corresponding optimal control law $u_\lambda$ satisfies
\[
\int_0^T \frac{r}{2}\|u_\lambda\|^2\mathrm{dt} \leq \int_0^T \frac{r}{2}\|u_0\|^2\mathrm{dt},
\] 
where $u_0$ is a continuous control law that transfers the state $y$ from
$y(0)$ to $p_\lambda$. $u_0$ is independent of $M$. 
We have
\[
y_\lambda(t) = \exp(At)\bar{x}_0+\int_0^t \exp(A(t-\sigma))Bu_\lambda(\sigma)
\mathrm{d\sigma}.
\]
Therefore,
\[\int_0^T\|y_\lambda\|^2\mathrm{dt}\leq K_1+ K_2 \int_0^T \|u_\lambda\|^2\mathrm{dt}+K_3 \Big(\int_0^T \|u_\lambda\|^2\mathrm{dt}\Big)^{\frac{1}{2}},
\]
for some positive constants $K_1,K_2,K_3$ which are independent of $M$.
Hence, $y_\lambda$ is uniformly bounded with $M$.

\subsection{Proof of Theorem \ref{theorem:fixed1}} 

The first point follows from Theorem \ref{theorem:fixedl} and (\ref{eq:index}). 
For 2) and 3), we define 
\[
a_N(\alpha)=\frac{N}{\xi_{12}(p_1-p_2)}\big(\beta_{12}^{T}x_{\alpha}^{0}-\delta_{12}-\theta_{12}\bar{x}_0-\xi_{12}p_2 \big ).
\]
We start by proving 2). 
Suppose that there does not exist any $\alpha$ in $\{0,...,N\}$ 
satisfying (\ref{eq:ineq}), (\ref{eq:ineq1}) or (\ref{eq:ineq2}). Zero does not satisfy (\ref{eq:ineq1}), 
hence $a_N(1)\leq 0 <1$. One does not satisfy (\ref{eq:ineq}) and $a_N(1)<1$, hence $a_N(2)\leq 1$. 
By induction, we have $a_N(N) \leq N-1$. Therefore, $N$ satisfies (\ref{eq:ineq2}). 
Thus, by contradiction, there exists $\alpha$ in $\{0,...,N\}$ satisfying (\ref{eq:ineq}), 
(\ref{eq:ineq1}) or (\ref{eq:ineq2}). We now prove the third point. Suppose that there exist 
multiple $\alpha$'s satisfying (\ref{eq:ineq}), (\ref{eq:ineq1}) or (\ref{eq:ineq2}). 
Let $\alpha_0$ be the least of these $\alpha$'s. If $\alpha_0<N$, then in view of 
$\xi_{12}(p_1-p_2) <0 $,  $a_N(\alpha_0+1)<\alpha_0 \leq a_N(\alpha_0)$. 
$a_N(j)$ is decreasing. Hence, for all 
$\alpha>\alpha_0$, $a_N(\alpha)\leq a_N(\alpha_0+1)<\alpha_0 <\alpha$. 
Therefore, $\alpha_0$ 
is the unique $\alpha$ satisfying (\ref{eq:ineq}), (\ref{eq:ineq1}) or (\ref{eq:ineq2}). 
If $\xi_{12}(p_1-p_2) <0 $, then the initial distribution for which $a_N(\alpha)$ is 
in $(0,1)$ for all $\alpha$ in $\{0,...,N\}$ does not have any $\alpha$ in $\{0,...,N\}$ 
satisfying (\ref{eq:ineq}), (\ref{eq:ineq1}) or (\ref{eq:ineq2}).


\section{}\label{Proof2}

\subsection{Proof of Theorem \ref{lemma:existance11}} 

We start by proving (i). Let $\hat{x}$ be a fixed point of $G_s$ 
and $\lambda_j = P_0(D_j(\hat{x}))$. 
By replacing the probabilities in the expression of $G_s$ by $\lambda_j$, $j=1,\dots,l$, we get $\hat{x}=G_s(\hat{x})=T_\lambda(\hat{x})$, 
where $\lambda=(\lambda_1,\dots,\lambda_l)$ and $T_\lambda$ are as defined 
in (\ref{eq:muapp1}). Hence, $\hat{x}$ is a fixed point of  $T_\lambda$. 
By Lemma \ref{lemma:existance1}, $\hat{x}(t)=R_1(t)\mu_0+R_2(t)p_\lambda$. 
By replacing this expression of $\hat{x}$ in 
$D_j(\hat{x})$, 
we get $\lambda=F_s(\lambda)$. Conversely, 
consider 
$\lambda=(\lambda_1,\dots,\lambda_l)$ in 
$\Delta_l$ such that 
$\lambda=F_s(\lambda)$ and let $\hat{x}(t)=R_1(t)\mu_0+R_2(t)p_\lambda$. The path 
$\hat{x}$ is the unique fixed point of $T_\lambda$ and
\[
\bigg(P_0(D_1(\hat{x})),\dots,P_0(D_l(\hat{x}))\bigg)=F_s(\lambda)=\lambda.
\]
Hence, $\hat{x}=T_\lambda(\hat{x})=G_s(\hat{x})$. 
We now prove the second point. Noting that the set $\Delta_l$ is convex and compact in 
$\mathbb{R}^l$, we just need to show that $F_s$ is continuous. 
Then, Brouwer's fixed point theorem \cite{FunAnaCon} ensures the existence 
of a fixed point. Let $\lambda_r$ be a sequence in $\Delta_l$ converging to $\lambda$. Let
\iftoggle{jou}{
\begin{multline*}
D_{kr}=\Big \{x\in \mathbb{R}^n \text{ such that}\\
(\beta_{kj})^Tx \leq \delta_{kj} + \theta_{kj}\mu_{0} + \xi_{kj}p_{\lambda_r},\, \forall j=1,\dots,l\Big\}
\end{multline*}
\begin{multline*}
D_{k}=\Big \{x\in \mathbb{R}^n \text{ such that}\\
(\beta_{kj})^Tx \leq \delta_{kj} + \theta_{kj}\mu_{0} + \xi_{kj}p_{\lambda},\, \forall j=1,\dots,l\Big\}.
\end{multline*}}{
\begin{equation*}
D_{kr}=\Big \{x\in \mathbb{R}^n \text{ such that }
(\beta_{kj})^Tx \leq \delta_{kj} + \theta_{kj}\mu_{0} + \xi_{kj}p_{\lambda_r},\, \forall j=1,\dots,l\Big\}
\end{equation*}
\begin{equation*}
D_{k}=\Big \{x\in \mathbb{R}^n \text{ such that }
(\beta_{kj})^Tx \leq \delta_{kj} + \theta_{kj}\mu_{0} + \xi_{kj}p_{\lambda},\, \forall j=1,\dots,l\Big\}.
\end{equation*}
}
We have
\iftoggle{jou}{
\begin{align*}
\Big | \big [F_s(\lambda_r)\big]_k-\big [F_s(\lambda)\big]_k \Big |&=
\bigg|\int_{\mathbb{R}^n}\! (1_{D_{kr}}(x)-1_{D_k}(x)) \, \mathrm{d}P_0(x) \bigg|\\
&\leq \int_{\mathbb{R}^n}\! \Big|1_{D_{kr}}(x)-1_{D_k}(x)\Big| \, \mathrm{d}P_0(x). 
\end{align*}}{
\begin{multline*}
\Big | \big [F_s(\lambda_r)\big]_k-\big [F_s(\lambda)\big]_k \Big |=
\bigg|\int_{\mathbb{R}^n}\! \Big(1_{D_{kr}}(x)-1_{D_k}(x)\Big) \, \mathrm{d}P_0(x) \bigg|
\leq \int_{\mathbb{R}^n}\! \Big|1_{D_{kr}}(x)-1_{D_k}(x)\Big| \, \mathrm{d}P_0(x). 
\end{multline*}
}
But, $D_{kr}$ and $D_{k}$ are regions delimited by hyperplanes. 
Hence, under Assumption \ref{assumption: agent spread},
\iftoggle{jou}{
\begin{multline*}
\int_{\mathbb{R}^n}\! \Big|1_{D_{kr}}(x)-1_{D_k}(x)\Big| \, \mathrm{d}P_0(x)=\\
\int_{\mathbb{R}^n}\! \Big|1_{\overset{\circ}{D}_{kr}}(x)-1_{\overset{\circ}{D}_k}(x)\Big| \, \mathrm{d}P_0(x). 
\end{multline*}}{
\begin{equation*}
\int_{\mathbb{R}^n}\! \Big|1_{D_{kr}}(x)-1_{D_k}(x)\Big| \, \mathrm{d}P_0(x)=
\int_{\mathbb{R}^n}\! \Big|1_{\overset{\circ}{D}_{kr}}(x)-1_{\overset{\circ}{D}_k}(x)\Big| \, \mathrm{d}P_0(x). 
\end{equation*}
}
But, 
$
\Big|1_{\overset{\circ}{D}_{kr}}(x)-1_{\overset{\circ}{D}_k}(x)\Big|\leq 2
$
and converges to zero for all $x$ in $\mathbb{R}^n$. Thus, by Lebesgue dominated 
convergence theorem \cite{Rudin87_Real}, the integral of this function converges 
to zero. This proves that $F_s$ is continuous. 
Finally, we prove (iii). For $l=2$, the fixed points of $F_s$ are of the 
form $(\alpha,1-\alpha)$. The set of the fixed points of $F_s$ is compact. Thus, the set of the first components of these fixed points is compact. Let $\alpha_0$ be the minimum of those first components. 
Consider $\alpha>\alpha_0$. Hence, \begin{multline*}
\Big \{(\beta_{12})^{T}x_i^{0}-\delta_{12}-\theta_{12}\mu_0-\xi_{12}p_2\leq 
\alpha \xi_{12}(p_1-p_2)\Big \}\\ \subset \{(\beta_{12})^{T}x_i^{0}-\delta_{12}-\theta_{12}\mu_0-\xi_{12}p_2\leq 
\alpha_0 \xi_{12}(p_1-p_2)\Big \},
\end{multline*}
 which implies
 \[\Big[F_s(\alpha,1-\alpha)\Big]_1\leq \Big[F_s(\alpha_0,1-\alpha_0)\Big]_1=\alpha_0<\alpha.\]
Thus, $(\alpha_0,1-\alpha_0)$ is the unique fixed point of $F_s$, and $\hat{x}(t)=R_1(t)\mu_0+R_2(t)p_{(\alpha_0,1-\alpha_0)}$ is the unique fixed point of $G_s$.

\subsection{Proof of Theorem \ref{Flat}}

We show in Theorem \ref{lemma:existance11} that the fixed points of $G_s$ can be one to one mapped to the fixed points of $F_s$. The initial states $x_i^0$ are distributed according to a Gaussian distribution $\mathcal{N}(\mu_0,\Sigma_0)$. Therefore, $\beta_{12}^Tx_i^0$ are distributed according to the normal distribution $\mathcal{N}\Big(\beta_{12}^T\mu_0,\beta_{12}^T\Sigma_0\beta_{12}\Big)$. Thus, one can analyze the dependence of $\big[F_s(\alpha,1-\alpha)\big]_1-\alpha$ on $\alpha$ to show that this function has a unique zero in $[0,1]$ in case $1)$ or $2)$ holds. Indeed, if 1) or 2) holds, the sign of the derivative with respect to $\alpha$ of $\big[F_s(\alpha,1-\alpha)\big]_1-\alpha$ does not change. Thus, this function is monotonic. This implies that $F_s$ and $G_s$ have unique fixed points.

\subsection{Proof of Theorem \ref{lemma:existanceinit}} 
We use Schauder's fixed point theorem \cite{FunAnaCon} to prove the existence 
of a fixed point. We start by showing that $G_p$ is a compact operator, that is 
continuous and maps bounded sets to relatively compact sets.
Let $\hat{x}$ be in $C([0,T],\mathbb{R}^n)$ and $\{\hat{x}_k\}_{k\in \mathbb{N}}$ 
be a sequence converging to $\hat{x}$ in $\big (C([0,T],\mathbb{R}^n),\|.\|_\infty\big)$. 
Let 
\iftoggle{jou}{
\begin{multline*}
Q_j=\max_{(\theta,t)\in \Theta \times [0,T]^2}\|\Phi_j^\theta(t)\|\\
+\max_{(\theta,t)\in \Theta \times [0,T]^4}\|\Psi_{j}^{\theta}(t)\|+\max_{\theta\in \Theta}\|M^\theta\|.
\end{multline*}}{
\begin{equation*}
Q_j > \max_{(\theta,t)\in \Theta \times [0,T]^2}\|\Phi_j^\theta(t)\|
+\max_{(\theta,t)\in \Theta \times [0,T]^4}\|\Psi_{j}^{\theta}(t)\|+\max_{\theta\in \Theta}\|M^\theta\|.
\end{equation*}
}
We have
\iftoggle{jou}{
\begin{multline*}
\|G_p(\hat{x}_k)-G_p(\hat{x})\|_\infty \leq \\ \sum_{j=1}^{l} Q_j \bigg \{ \frac{qT^2}{r} \|\hat{x}_k-\hat{x}\|_\infty + 
 V_{1j} + 
  \frac{Q_j\|p_j\|T+q\|\hat{x}\|_\infty T^2}{r}V_{2j} \bigg \}, 
\end{multline*}}{
\begin{equation*}
\|G_p(\hat{x}_k)-G_p(\hat{x})\|_\infty \leq  \sum_{j=1}^{l} Q_j \bigg \{ \frac{qT^2}{r} \|\hat{x}_k-\hat{x}\|_\infty + 
 V_{1j} + 
  \frac{Q_j\|p_j\|T+q\|\hat{x}\|_\infty T^2}{r}V_{2j} \bigg \}, 
\end{equation*}
}
where
\begin{align*}
& V_{1j}=\int_{\Theta} \int_{\mathbb{R}^n}  
\Big|1_{D^{\theta}_{j}(\hat{x}_k)}(x^0)-1_{D^{\theta}_{j}(\hat{x})}(x^0)\Big|\|x^0\| \, \mathrm{d} P_0\mathrm{d} P_\theta\\
& V_{2j}=\int_{\Theta} \int_{\mathbb{R}^n}\Big|1_{D^{\theta}_{j}(\hat{x}_k)}(x^0)-1_{D^{\theta}_{j}(\hat{x})}(x^0)\Big| \, \mathrm{d} P_0\mathrm{d} P_\theta.
\end{align*} 
Under Assumption \ref{assumption: measure support},
\begin{align*}
& V_{1j}=\int_{\Theta} \int_{\mathbb{R}^n}\Big|1_{\overset{\circ}{D}^{\theta}_{j}(\hat{x}_k)}(x^0)-1_{\overset{\circ}{D}^{\theta}_{j}(\hat{x})}(x^0)\Big|\|x^0\| \, \mathrm{d} P_0\mathrm{d} P_\theta.
\end{align*} 
But, 
\[\Big|1_{\overset{\circ}{D}^{\theta}_{j}(\hat{x}_k)}(x^0)-1_{\overset{\circ}{D}^{\theta}_{j}(\hat{x})}(x^0)\Big|\|x^0\|\leq 2\|x^0\|\]
and converges to zero for all $(x^0,\theta)$ in $\mathbb{R}^n \times \Theta$. 
We have $\mathbb{E}\|x^0\| < \infty$. Therefore, by Lebesgue's dominated convergence theorem \cite{Rudin87_Real}, $V_{1j}$ converges to zero.    
By the same technique, we prove that $V_{2j}$ converges to zero. Hence, $G_p$ is continuous. 
Let $V$ be a bounded subset of $C([0,T],\mathbb{R}^n)$. 
Let $\{G_p(\hat{x}_k)\}_{k\in \mathbb{N}} \in G_p(V)$.
By the continuity of $\Phi_j^\theta(\sigma,t)$ with respect to $(\sigma,t,\theta)$, 
of its derivative with respect to $t$ and $\sigma$, and by the boundedness of $\hat{x}_k$, 
one can prove that for all $(t, s)$ in $[0,T]^2$, 
\begin{equation*}
\|G_p(\hat{x}_k)(t)-G_p(\hat{x}_k)(s)\|\leq \Big (K_1 \mathbb{E}\|x^0\| + K_2 \Big ) |t-s|, 
\end{equation*}
where $K_1$ and $K_2$ are positive constants.   
This inequality implies the uniform boundedness and equicontinuity of 
$\{G_p(\hat{x}_k)\}_{k\in \mathbb{N}}$. By Arzela-Ascoli Theorem \cite{FunAnaCon}, 
there exists a convergent subsequence of $\{G_p(\hat{x}_k)\}_{k\in \mathbb{N}}$. 
Hence, $G_p(V)$ and its closure are compact sets, and $G_p$ is a compact operator. 
%
Now, we construct a nonempty, bounded, closed, convex subset 
$U \subset C([0,T],\mathbb{R}^n)$ such that $G_p(U)\subset U$. Let
$Q=\max(k_1+k_2,k_3)$, where $k_1$, $k_2$ and $k_3$ are defined in
(\ref{bounds}). 
We start by defining on $[0,T]$ the 
function 
$R(t)=Q\cos (\sqrt{Q}t)+Q\tan(\sqrt{Q}T)
\sin (\sqrt{Q}t)$. Under Assumption \ref{assumption: bounds}, $R(t)$ is positive. Moreover, $R$ 
satisfies $R(t)= Q+Q\int_{0}^{t} \; \int^{T}_{\sigma} \; R(\tau) \, \mathrm{d}\tau \mathrm{d}\sigma$. 
Let 
\[U=\Big \{x\in C([0,T],\mathbb{R}^n)|\, \|x(t)\| \leq R(t), \, \forall t\in [0,T]\Big \}.\]
The set $U$ is an nonempty, bounded, closed and convex subset of $C([0,T],\mathbb{R}^n)$. For all $x \in U$, $\forall t \in [0,T]$,
\begin{equation*}
\|G_p(x)(t)\| \leq Q+Q\int_{0}^{t} \; \int^{T}_{\sigma} \; R(\tau) \, \mathrm{d}\tau \mathrm{d}\sigma = R(t).
\end{equation*}
Hence, $G_p(U)\subset U$. By Schauder's Theorem, $G_p$ has a fixed point in $U$.

\section{Proof of Theorem \ref{theorem:nash}} \label{Proof3}
We consider an arbitrary agent $i\in \{1,...,N\}$ applying an arbitrary full state feedback control 
law $u_i$. Suppose that this agent $i$ can profit by a unilateral 
deviation from the decentralized strategies. This means that
\begin{equation} \label{maxprofit}
 J_{i} (u_i, u^*_{-i})
\leq J_{i} (u_i^*, u^*_{-i}).
\end{equation}
In the following, we prove that this profit is bounded by $\epsilon$.
We denote respectively 
by $x_i$ and $x^*_j$ the states 
corresponding to $u_i$ and $u^*_j$.
In view of (\ref{eq:sysininitial}), the compactness of $\Theta$, the continuity of $x_j^*$ with respect 
to $\theta$ and $\mathbb{E}\|x_i^0\|^2 < \infty$, the right hand side 
of (\ref{maxprofit}) is bounded by $Q_1$ independently of $N$.
For any $X$ and $Y$ in $C([0,T],\mathbb{R}^n)$, we define 
\[<X|Y>=\mathbb{E}\bigg(\int_{0}^{T} \, X^T(t)Y(t) \, \mathrm{d}t\Big | x_i^0\bigg )\] 
and $\|X\|_2=\sqrt{<X|X>}$.
We have
\iftoggle{jou}{
\begin{multline*}
J_{i}(u_i, u^*_{-i})=
 J_{i}\Big(x_i,\hat{x},x_i^0\Big)+ \frac{q}{2} \Big \| \hat{x}-\frac{1}{N}\sum_{j=1}^{N} x_{j}^* \Big \|^{2}_2 \\ 
+\frac{q}{2N^2} \| x_{i}^*-x_{i} \|^{2}_2 +S_1+S_2+S_3,
\end{multline*}}{
\begin{multline*}
J_{i}(u_i, u^*_{-i})=
 J_{i}\Big(x_i(u_{i}),\hat{x},x_i^0\Big)+ \frac{q}{2} \Big \| \hat{x}-\frac{1}{N}\sum_{j=1}^{N} x_{j}^* \Big \|^{2}_2
+\frac{q}{2N^2} \| x_{i}^*-x_{i} \|^{2}_2 +S_1+S_2+S_3,
\end{multline*}
}
where
\begin{align*}
& S_1= \frac{q}{N}\Big <  x_{i}^*-x_{i} \Big |  x_{i}-\hat{x} \Big > \\
& S_2= \frac{q}{N}\Big <  x_{i}^*-x_{i} \Big |  \hat{x}-\frac{1}{N}\sum_{j=1}^{N} x_{j}^* \Big > \\
& S_3= q\Big <  \hat{x}-\frac{1}{N}\sum_{j=1}^{N} x_{j}^* \Big | x_{i}-\hat{x} \Big >,
\end{align*}
with $\hat{x}$ is a fixed point of $G_p$.
By the Cauchy-Schwarz inequality,
\begin{align*}
|S_1| \leq \frac{q}{N}\Big \|  x_{i}^*-x_{i} \Big \|_2 \Big \|  x_{i}-\hat{x} \Big \|_2.
\end{align*}
In view of (\ref{maxprofit}) and the bound $Q_1$, $\Big \|  x_{i}^*-x_{i} \Big \|_2$ and $\Big \|  x_{i}-\hat{x} \Big \|_2$ are bounded. Thus, $|S_1 |\leq \eta_1/N$, where $\eta_1>0$. Similarly, $|S_2 |\leq \eta_2/N$, where $\eta_2>0$. 
We define 
\begin{equation*}
\alpha_N=\Big \| \hat{x}-\frac{1}{N}\sum_{j=1}^{N} \mathbb{E}x_{j}^* \Big \|_2=
\Big \| \int_\Theta \, \bar{x}^{\theta} \, \mathrm{d}P_\theta -  \int_\Theta  \, \bar{x}^{\theta} \, \mathrm{d}P_\theta^N \Big \|_2, 
\end{equation*}

where $\bar{x}^{\theta}$ is defined in (\ref{meaninit}).
We have
\begin{equation*}
\Big \| \hat{x}-\frac{1}{N}\sum_{j=1}^{N} x_{j}^* \Big \|^{2}_2 \leq  2\alpha_N^2 + 2\Big \| \frac{1}{N}\sum_{j=1}^{N} \Big(\mathbb{E}x_{j}^*- x_{j}^*\Big) \Big \|^{2}_2.
\end{equation*}
By the compactness of $[0,T]\times \Theta $, the family of functions $\bar{x}^{\theta}(t)$ defined on $\Theta$ and indexed by $t$ is uniformly bounded and equicontinuous. By Corollary $1.1.5$ of \cite{stroock1979multidimensional}, we deduce
\begin{align*}
\lim\limits_{N \rightarrow +\infty} \sup\limits_{t \in [0,T]}\Big \| \hat{x}(t)-\frac{1}{N}\sum_{j=1}^{N} \mathbb{E}x_{j}^*(t) \Big \|=0. 
\end{align*}
Thus, $\alpha_N$ converges to $0$ as $N$ increases to infinity.
By the independence of the initial conditions (and thus the independence of $x_j^*$, $j=1,\dots,N$) and the assumption $\mathbb{E}\|x_{i}^0\|^2<\infty$,  we deduce that
\[
\Big \| \frac{1}{N}\sum_{j=1}^{N} \Big(\mathbb{E}x_{j}^*- x_{j}^*\Big) \Big \|^{2}_2 = O(1/N).
\]
Thus, $S_3$ and $\Big|J_{i}\Big(x_i^*,\hat{x},x_i^0\Big)-J_{i}\Big(u_i^*, u^*_{-i}\Big)\Big|$ converge to $0$ as $N$ increases to infinity.
By optimality, we have $J_{i}\Big(x_i^*,\hat{x},x_i^0\Big)\leq J_{i}\Big(x_i,\hat{x},x_i^0\Big)$. Therefore,  $
J_{i}(u_i, u^*_{-i}) \geq J_{i}(u_i^*, u^*_{-i})+\epsilon_N
$,
where $\epsilon_N=J_{i}\Big(x_i^*,\hat{x},x_i^0\Big)-J_{i}\Big(u_i^*, u^*_{-i}\Big)+S_1+S_2+S_3$ converges to $0$ as $N$ increases to infinity.

\ifCLASSOPTIONcaptionsoff
  \newpage
\fi

\bibliographystyle{IEEEtran}
\bibliography{IEEEabrv,mfg}

\begin{thebibliography}{10}
\providecommand{\url}[1]{#1}
\csname url@samestyle\endcsname
\providecommand{\newblock}{\relax}
\providecommand{\bibinfo}[2]{#2}
\providecommand{\BIBentrySTDinterwordspacing}{\spaceskip=0pt\relax}
\providecommand{\BIBentryALTinterwordstretchfactor}{4}
\providecommand{\BIBentryALTinterwordspacing}{\spaceskip=\fontdimen2\font plus
\BIBentryALTinterwordstretchfactor\fontdimen3\font minus
  \fontdimen4\font\relax}
\providecommand{\BIBforeignlanguage}[2]{{%
\expandafter\ifx\csname l@#1\endcsname\relax
\typeout{** WARNING: IEEEtran.bst: No hyphenation pattern has been}%
\typeout{** loaded for the language `#1'. Using the pattern for}%
\typeout{** the default language instead.}%
\else
\language=\csname l@#1\endcsname
\fi
#2}}
\providecommand{\BIBdecl}{\relax}
\BIBdecl

\bibitem{Rab2014}
R.~Salhab, R.~P. Malham\'e, and J.~Le~Ny, ``Consensus and disagreement in
  collective homing problems: A mean field games formulation,'' in
  \emph{Proceedings of the 53rd IEEE Conference on Decision and Control}, Dec
  2014, pp. 916--921.

\bibitem{Rab2015}
------, ``A dynamic game model of collective choice in multi-agent systems,''
  in \emph{Proceedings of the 54th IEEE Conference on Decision and Control},
  dec 2015.

\bibitem{Leonard12_decision}
N.~E. Leonard, T.~Shen, B.~Nabet, L.~Scardovi, I.~D. Couzin, and S.~A. Levin,
  ``Decision versus compromise for animal groups in motion,'' \emph{Proceedings
  of the National Academy of Sciences}, vol. 109, no.~1, pp. 227--232, 2012.

\bibitem{Couzin05_leadership}
I.~D. Couzin, J.~Krause, N.~R. Franks, and S.~A. Levin, ``Effective leadership
  and decision-making in animal groups on the move,'' \emph{Nature}, vol. 433,
  pp. 513--516, 2005.

\bibitem{Merill99_vote}
S.~Merrill and B.~Grofman, \emph{A Unified Theory of Voting: Directional and
  Proximity Spatial Models}.\hskip 1em plus 0.5em minus 0.4em\relax Cambridge
  University Press, 1999.

\bibitem{Seeley91_honeyBees}
T.~D. Seeley, S.~Camazine, and J.~Sneyd, ``Collective decision-making in honey
  bees: how colonies choose among nectar sources,'' \emph{Behavioral Ecology
  and Sociobiology}, vol.~28, pp. 277--290, 1991.

\bibitem{Camazin99_honeyBees}
S.~Camazine, P.~K. Visscher, J.~Finley, and R.~S. Vetter, ``House-hunting by
  honey bee swarms: collective decisions and individual behaviors,''
  \emph{Insectes Sociaux}, vol.~46, no.~4, pp. 348--360, November 1999.

\bibitem{Tien04_fishSchools}
J.~H. Tien, S.~A. Levin, and D.~I. Rubenstein, ``Dynamics of fish schools:
  identifying key decision rules,'' \emph{Evolutionary Ecology Research},
  vol.~6, pp. 555--565, 2004.

\bibitem{Aoki_fishSimulation}
I.~Aoki, ``A simulation study on the schooling mechanism in fish,''
  \emph{Bulletin of the Japanese Society for the Science of Fish}, vol.~48, pp.
  1081--1088, 1982.

\bibitem{Pratt06_algoDecisionMaking}
S.~C. Pratt and D.~J.~T. Sumpter, ``A tunable algorithm for collective
  decision-making,'' \emph{Proceedings of the National Academy of Sciences},
  vol. 103, pp. 15\,906--15\,910, 2006.

\bibitem{Daron13_op}
D.~Acemoglu and A.~Ozdaglar, ``Opinion dynamics and learning in social
  networks,'' \emph{Dynamic Games and Applications}, vol. 1.1, pp. 3--49, 2010.

\bibitem{Hegselmann02_op}
R.~Hegselmann and U.~Krause, ``Opinion dynamics and bounded confidence models,
  analysis, and simulation,'' \emph{Journal of Artifical Societies and Social
  Simulation}, vol.~5, 2002.

\bibitem{Whittle70_homing}
P.~Whittle and P.~Gait, ``Reduction of a class of stochastic control
  problems,'' \emph{IMA Journal of Applied Mathematics}, vol.~6, no.~2, pp.
  131--140, 1970.

\bibitem{Whittle82}
P.~Whittle, \emph{Optimization over time}.\hskip 1em plus 0.5em minus
  0.4em\relax Wiley, 1982.

\bibitem{Kuhn85_homing}
J.~Kuhn, ``The risk-sensitive homing problem,'' \emph{Journal of Applied
  Probability}, vol.~22, pp. 796--803, 1985.

\bibitem{Lef04}
M.~Lefebvre, ``A homing problem for diffusion processes with control-dependent
  variance,'' \emph{The Annals of Applied Probability}, vol.~14, pp. 786--795,
  2004.

\bibitem{Lef_GLQG}
M.~Lefebvre and F.~Zitouni, ``General {LQG} homing problems in one dimension,''
  \emph{International Journal of Stochastic Analysis}, vol. 2012, 2012.

\bibitem{Nourian11_collective}
N.~Nourian, R.~P. Malham\'e, M.~Huang, and P.~E. Caines, ``Mean-field {NCE}
  formulation of estimation-based leader-follower collective dynamics,''
  \emph{International Journal of Robotics and Automation}, vol.~26, no.~1, pp.
  120--129, 2011.

\bibitem{Mesbah10_graphth}
M.~Mesbahi and M.~Egerstedt, \emph{Graph Theoretic Methods in Multiagent
  Networks}, First, Ed.\hskip 1em plus 0.5em minus 0.4em\relax Princeton
  University Press, 2010.

\bibitem{DistCtrlRobotNetw}
F.~Bullo, J.~Cort\'es, and S.~Mart{\'\i}nez, \emph{Distributed Control of
  Robotic Networks}, ser. Applied Mathematics Series.\hskip 1em plus 0.5em
  minus 0.4em\relax Princeton University Press, 2009.

\bibitem{jerome13_ada}
J.~{Le Ny} and G.~J. Pappas, ``Adaptive deployment of mobile robotic
  networks,'' \emph{IEEE Transactions on automatic control}, vol.~58, pp.
  654--666, 2013.

\bibitem{Kappelman2005}
F.~Koppelman and V.~Sathi, ``Incorporating variance and covariance
  heterogeneity in the generalized nested logit model: an application to
  modeling long distance travel choice behavior,'' \emph{Transportation
  Research}, vol.~39, pp. 825--853, 2005.

\bibitem{Bhat2004}
C.~Bhat and J.~Guo, ``A mixed spatially correlated logit model: formulation and
  application to residential choice modeling,'' \emph{Transportation Research},
  vol.~38, pp. 147--168, 2004.

\bibitem{burke2003physician}
M.~A. Burke, G.~Fournier, and K.~Prasad, \emph{Physician social networks and
  geographical variation in medical care}.\hskip 1em plus 0.5em minus
  0.4em\relax Center on Social and Economic Dynamics, 2003.

\bibitem{Brock2001}
W.~Brock and S.~Durlauf, ``Discrete choice with social interactions,''
  \emph{Review of Economic Studies}, pp. 147--168, 2001.

\bibitem{Huang07_Large}
M.~Huang, P.~E. Caines, and R.~P. Malham\'e, ``Large-population cost-coupled
  {LQG} problems with nonuniform agents: Individual-mass behavior and
  decentralized epsilon-{N}ash equilibria,'' \emph{IEEE Transactions on
  Automatic Control}, vol.~52, no.~9, pp. 1560--1571, 2007.

\bibitem{Lasry07_MFG}
J.~M. Lasry and P.~L. Lions, ``Mean field games,'' \emph{Japanese Journal of
  Mathematics}, vol.~2, pp. 229--260, 2007.

\bibitem{Huang03_thesis}
M.~Huang, ``Stochastic control for distributed systems with applications to
  wireless communications,'' Ph.D. dissertation, McGill University, 2003.

\bibitem{Huang03_wirelessPower}
M.~Huang, P.~E. Caines, and R.~P. Malham\'e, ``Individual and mass behaviour in
  large population stochastic wireless power control problems: centralized and
  {N}ash equilibrium solutions,'' in \emph{Proceedings of the 42nd IEEE
  Conference on Decision and Control}, Maui, Hawaii, 2003, pp. 98--103.

\bibitem{Huang06_particles}
M.~Huang, R.~P. Malham\'e, and P.~E. Caines, ``Nash certainty equivalence in
  large population stochastic dynamic games: Connections with the physics of
  interacting particle systems,'' in \emph{Proceedings of the 44th IEEE
  Conference on Decision and Control}, San Diego, CA, 2006, pp. 4921--4926.

\bibitem{huang2006large}
------, ``Large population stochastic dynamic games: closed-loop
  {McKean-Vlasov} systems and the {Nash} certainty equivalence principle,''
  \emph{Communications in Information \& Systems}, vol.~6, no.~3, pp. 221--252,
  2006.

\bibitem{lasry2006jeux}
J.~M. Lasry and P.~L. Lions, ``Jeux {\`a} champ moyen. {I}--le cas
  stationnaire,'' \emph{Comptes Rendus Math{\'e}matique}, vol. 343, no.~9, pp.
  619--625, 2006.

\bibitem{lasry2006jeux2}
------, ``Jeux {\`a} champ moyen. {II}--horizon fini et contr{\^o}le optimal,''
  \emph{Comptes Rendus Math{\'e}matique}, vol. 343, no.~10, pp. 679--684, 2006.

\bibitem{FunAnaCon}
J.~B. Conway, \emph{A Course in Functional Analysis}, ser. Graduate Texts in
  Mathematics.\hskip 1em plus 0.5em minus 0.4em\relax Springer-Verlag, 1985.

\bibitem{anderson2007optimal}
B.~D. Anderson and J.~B. Moore, \emph{Optimal control: linear quadratic
  methods}.\hskip 1em plus 0.5em minus 0.4em\relax Dover Publications, 2007.

\bibitem{Rugh1993}
W.~Rugh, \emph{Linear System Theory}, ser. Prentice-Hall information and
  systems sciences series.\hskip 1em plus 0.5em minus 0.4em\relax Prentice
  Hall, 1993.

\bibitem{broyden1965class}
C.~G. Broyden, ``A class of methods for solving nonlinear simultaneous
  equations,'' \emph{Mathematics of computation}, pp. 577--593, 1965.

\bibitem{nakajima2007measuring}
R.~Nakajima, ``Measuring peer effects on youth smoking behaviour,'' \emph{The
  Review of Economic Studies}, vol.~74, no.~3, pp. 897--935, 2007.

\bibitem{huang2012social}
M.~Huang, P.~E. Caines, and R.~P. Malham\'e, ``Social optima in mean field
  {LQG} control: centralized and decentralized strategies,'' \emph{Automatic
  Control, IEEE Transactions on}, vol.~57, no.~7, pp. 1736--1751, 2012.

\bibitem{Rudin87_Real}
W.~Rudin, \emph{Real and Complex Analysis}, 3rd~ed.\hskip 1em plus 0.5em minus
  0.4em\relax McGraw-Hill Inc., 1987.

\bibitem{achdou2010mean}
Y.~Achdou and I.~Capuzzo-Dolcetta, ``Mean field games: Numerical methods,''
  \emph{SIAM Journal on Numerical Analysis}, vol.~48, no.~3, pp. 1136--1162,
  2010.

\bibitem{stroock1979multidimensional}
D.~W. Stroock and S.~S. Varadhan, \emph{Multidimensional diffussion
  processes}.\hskip 1em plus 0.5em minus 0.4em\relax Springer Science \&
  Business Media, 1979, vol. 233.

\end{thebibliography}

\end{document}